\documentclass[journal,12pt,onecolumn,draftclsnofoot]{IEEEtran}

\usepackage[utf8]{inputenc}
\usepackage{cite}

\usepackage[cmex10]{amsmath}
\usepackage{amssymb,latexsym,amsthm}
\usepackage{epsfig,epsf,epstopdf,subfigure}
\usepackage{caption}
\usepackage{color}
\usepackage{graphicx}
\usepackage{cool}
\usepackage{bbm}

\usepackage{pst-plot}
\usepackage{pstricks-add}
\usepackage{breqn}
\usepackage{algorithm,algorithmic}
\usepackage{psfrag}
\newcommand{\newoperator}[3]{\newcommand*{#1}{\mathop{#2}#3}}

\newoperator{\tr}{\mathrm{tr}}{\nolimits}
\newoperator{\diag}{\mathrm{diag}}{\nolimits}
\newoperator{\rank}{\mathrm{rank}}{\nolimits}
\newoperator{\myperm}{\mathrm{perm}}{\nolimits}

\renewcommand{\vec}[1]{\boldsymbol{#1}}

\newtheorem{lem}{\bf{Lemma}}
\newtheorem{prop}{\bf{Proposition}}
\newtheorem{rem}{\bf{Remark}}

\newtheorem{assumption}{\bf{Assumption}}

\newenvironment{lemma}
{\par\noindent \lem \begin{itshape}\noindent}
{\end{itshape}}

\newcommand{\tabincell}[2]{\begin{tabular}{@{}#1@{}}#2\end{tabular}}
\usepackage{mathtools} 
\title{\Large{Wirelessly-powered Sensor Networks:\\ Power Allocation for Channel Estimation and Energy Beamforming}}
\begin{document}
\author{\IEEEauthorblockN{Rong~Du,~Hossein Shokri Ghadikolaei,~Carlo~Fischione}\thanks{The paper has been accepted in IEEE Transactions on Wireless Communications on Jan. 19th, 2020}\\
		\IEEEauthorblockA{School of Electrical Engineering and Computer Science}\\
        \IEEEauthorblockA{KTH Royal Institute of Technology, Stockholm, Sweden}\\
		\IEEEauthorblockA{Email: \{rongd, hshokri, carlofi\}@kth.se} \\
	}
	\maketitle
\begin{abstract}
Wirelessly-powered sensor networks (WPSNs) are becoming increasingly important in different monitoring applications. We consider a WPSN where a multiple-antenna base station, which is dedicated for energy transmission, sends pilot signals to estimate the channel state information and consequently shapes the energy beams toward the sensor nodes. Given a fixed energy budget at the base station, in this paper, we investigate the novel problem of optimally allocating the power for the channel estimation and for the energy transmission. We formulate this non-convex optimization problem for general channel estimation and beamforming schemes that satisfy some qualification conditions. We provide a new solution approach and a performance analysis in terms of optimality and complexity. We also present a closed-form solution for the case where the channels are estimated based on a least square channel estimation and a maximum ratio transmit beamforming scheme. The analysis and simulations indicate a significant gain in terms of the network sensing rate, compared to the fixed power allocation, and the importance of improving the channel estimation efficiency.
\end{abstract}

\begin{IEEEkeywords}
Wirelessly-powered sensor network, wireless energy transfer, power allocation, channel acquisition, non-linear energy harvesting
\end{IEEEkeywords}

\section{Introduction}

Traditional battery-powered wireless sensor networks suffer a major problem of the limited energy budget of the nodes. Unless the battery of the nodes are replaced periodically, the network lifetime is limited. Thus, for long-term monitoring applications, rechargeable sensor networks are more appealing  than traditional battery-powered ones~\cite{sudevalayam2011energy}.
A promising technique to recharge the sensor nodes is called wireless energy transmission (WET)~\cite{xie2013wireless,zeng2017communications}, in which electromagnetic waves periodically recharge sensor nodes to extend their lifetime. Compared to the ambient energy harvesting, WET provides a better predictability, controllability, and reliability \cite{lu2015wireless}, leading to a more consistent performance of the network.

\subsection{Related Works and Motivations}

WET systems can be broadly divided into two categories, according to the transmission of the data: simultaneous wireless information and power transfer~\cite{zhong2015wireless} and wirelessly-powered communication networks (WPCNs)~\cite{bi2015wireless}. In the first category, the transmitter simultaneously sends energy and data at the same time, and the receiver allocates some resources (e.g., time, power, or antennas) for energy harvesting and the rest for data communication~\cite{krikidis2014simultaneous}. In the second category,  WPCN~\cite{bi2015wireless,bi2016wireless}, the process of energy and data transmissions is sequential, meaning that the energy receivers send their data using the energy harvested from the transmitters. Most of the existing studies in WPCN focus on maximizing the throughput of wireless devices via optimizing frequency or time schedules for energy transmissions and for data transmissions~\cite{sun2014joint,yang2015throughput,lee2016sum,bi2016wireless}.

\begin{figure}[t]
	\centering
	\includegraphics[width=0.4\textwidth]{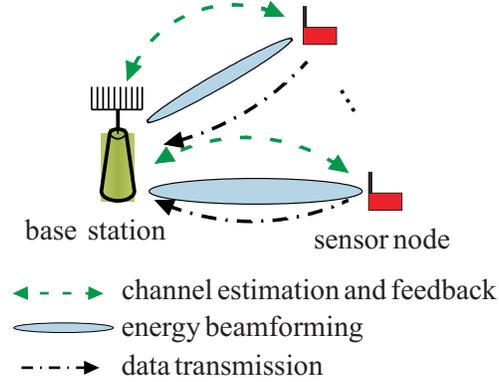}
	\caption{The wirelessly-powered sensor network considered in this paper. The network consists of one  base station and multiple sensor nodes. The base station uses energy beamforming to transmit energy to the nodes, and the nodes use the harvested energy for sensing and data transmission back to a sink node.} 
	\vspace{-0.8cm}
	\label{Fig:NetSys}
\end{figure}

The severe propagation loss in wireless medium may result in a small received energy that may not be enough for the data transmission task~\cite{xie2013wireless}. To overcome this problem, instead of adopting an arbitrarily large transmission power, which is not possible due to the safety issue~\cite{sohul2015spectrum,galinina2016feasibility,alsaba2018beamforming}, we can substantially improve the WET efficiency by energy beamforming~\cite{liu2014multi,lee2017distributed}. More specifically, we can steer the energy toward the receivers, such that with the same transmission power the receivers can harvest substantially more energy compared to an omnidirectional energy transmission scheme.  
To this end, we need multiple antennas and channel state information (CSI). Fig.~\ref{Fig:NetSys} illustrates a typical wirelessly-powered sensor networks (WPSNs)~\cite{choi2017wireless}, which is a special case of WPCN. In WPSNs, some energy sources, hereafter called base stations (BSs), provide energy to the nodes using WET, and the nodes use the received energy to make measurements and send them to a sink, which could be an energy source. In this paper, we focus on the WPSNs with one BS.

There are some results on optimal beamforming design for energy transfer. With perfect CSI, the work in~\cite{xu2016general} shows that the optimal energy beamforming in terms of received energy of a point-to-point MIMO system can be achieved by the eigenvector corresponding to the largest eigenvalue of the channel matrix. For a WPCN with multiple energy receivers, the authors of~\cite{sun2014joint} study a joint time allocation and energy beamforming problem to maximize the network sum-throughput and provide a solution approach based on semi-definite relaxation. In a wireless sensor network, however, monitoring performance and lifetime are more important than the network sum-rate.
References~\cite{Du2017ICC,xie2012making,xie2015mobile,Du2016Lifetime} study the  monitoring performance and the lifetime of WPSNs, assuming that the energy sources (BSs) know perfect CSI a priori  at no cost. 
However, in practice, the energy transmitters should always spend power and time to acquire CSI~\cite{bi2015wireless}, and the WET efficiency greatly depends on the channel estimation quality. 

The accuracy of the CSI estimation for WET has thus been investigated in~\cite{kashyap2016feasibility,zeng2015optimized,zeng2017communications,lee2017distributed,abeywickrama2018wireless,abeywickrama2018cluster,xu2014energy,huang2018waveform}. In particular, the work in~\cite{kashyap2016feasibility} investigates the interplay between the power allocation for channel estimation  and the expected received energy at the receiver. However, in the context of wireless sensor networks, there is no study for  the power allocation for channel estimation and energy transmission to maximize the network performance. 
The authors of~\cite{le2018joint} study a network with one multiple-antenna energy transmitter and one single-antenna energy receiver, and formulate an optimization problem on how much power and time to spend in channel estimation and channel feedback to maximize the rate in downlink data transmission. The objective function is non-convex and not analytically tractable. Thus, they maximize an upper bound of the downlink data rate instead.
The work in~\cite{abeywickrama2018wireless} considers a similar setting, where the transmitter sends predefined energy beams to the receiver, and the receiver feeds back the strength of the received signal. The authors optimize the size of the training codebook for channel acquisition to balance the trade-off between channel estimation precision and energy transmission time. These results are extended to multiple single-antenna energy receivers in~\cite{abeywickrama2018cluster}. Based on the feedback, the transmitter employs a maximum likelihood channel estimation method and clusters the receivers according to their channel phases during energy transmission. The work in~\cite{lee2016energy} studies a network with one single-antenna energy transmitter and multiple single-antenna energy receivers, where the transmitter applies minimum mean square error (MMSE) to estimate the channels. The authors optimize the time allocation for channel estimation and energy transmission to maximize the energy efficiency of the network.  In~\cite{xu2014energy}, the authors investigate energy harvesting (EH) with an one-bit feedback, where every receiver only reports whether the harvested energy in the current time interval is larger than that of the previous interval, and the transmitter estimates the channel using analytic center cutting plane method. With a similar training scheme, the authors of~\cite{huang2018waveform} consider the cases where the energy transmitter sends multi-sine energy signals over different frequencies, and model the EH by a non-linear function.
Reference~\cite{zeng2015optimized} proposes a pilot design approach that transmits pilots from the energy receiver to the energy transmitter for a point-to-point WET system and trades-off the channel estimation accuracy and the corresponding spent energy. Although this work provides the optimal solution for some special cases, the general cases remain unsolved due to the intractability of the expressions. Reference~\cite{lee2017distributed} extends this work to a scenario with multiple transmitters and single receiver. 

Most of the above-mentioned studies use a linear EH model in the formulation. However, in reality, the EH is non-linear due to circuit sensitivity limitations, current leakage~\cite{boshkovska2015practical}, and filters in the circuits~\cite{clerckx2018wireless}, among others. A linear EH model sometimes can be considered as a special case of the non-linear harvesting models~\cite{clerckx2018wireless}. Therefore, the algorithms designed for the problem with linear EH models and the resulting insights  may not hold for a general non-linear model. There are some works studying robust energy transmission with non-linear EH model~\cite{boshkovska2017robust,lu2018coordinated}. They model the imperfect CSI by a channel error with either a deterministic bound, or a random one with a deterministic and bounded variance. However, they do not study how to improve the EH performance by properly allocating resources in channel acquisition. Moreover, recent attempts on the design of algorithms for the non-linear EH cases are based on some restrictive assumptions~\cite{boshkovska2015practical,lu2018coordinated}. For example, the solution for~\cite{boshkovska2015practical} works for a specific EH model, and whether it is valid for other models is unsure. The approach in~\cite{lu2018coordinated} can deal with the cases where the EH related constraint requires that the harvested energy is larger or smaller than a pre-defined threshold, which allows us to transform the EH related constraint  into a linear constraint on the received energy. The solutions for a fairly general EH model are still open. In this paper, we try to address this important research gap.  We consider a wireless sensor network with multiple energy receivers, where the input-output model of the EH circuits of these energy receivers does not need to be a particular function, but rather whatever function as long as it is monotone and increasing. Hereafter, we refer to this EH model (formally defined in Section II) as a generic model to distinguish it from a particular function/model. We investigate the fundamental trade-off between the channel estimation and energy transmission performance in the presence of an arbitrary approach for pilot transmission, channel estimation, and beamforming. We exemplify the use of our framework and show its application in benchmarking the performance of various approaches for pilot transmission, channel estimation, and beamforming. Our novelty is the joint consideration of the following: 1) controlling the CSI quality to benefit the energy transmission process, instead of having it as a given input; 2) generic EH model and channel estimation methods.
\subsection{Contributions}	
In this paper, we consider a WPSN comprised of one BS and multiple energy receivers, and investigate a new problem of optimal power allocation for channel estimation and energy transmission that maintains a required monitoring performance throughout the network. We substantially extend our preliminary results~\cite{rong2018power} by optimizing the power allocation for a generic model that characterizes the performance of the channel estimation and energy beamforming schemes. We also extend our preliminary results in~\cite{rong2018power}  to a class of non-linear EH models as long as the function that characterizes the input-output relation of the energy of the EH circuit is monotone and increasing. We develop a novel solution approach based on a bisection search and iterative feasibility checking. We exemplify the proposed solution approach for a specific case of least square (LS) channel estimation and maximum ratio transmission for energy beamforming.
To summarize, the main contributions of the paper are as follows:


\begin{itemize}
	\item We propose a novel problem of power allocation for channel estimation and energy transmission for multiple sensor nodes, to maximize the monitoring performance of a WPSN. We consider a generic EH model, which makes our optimization problem more challenging. However, due to the generality in the model, our results are general and we can directly apply them to any channel estimation and energy beamforming schemes that satisfy the technical conditions in Section~\ref{sec:CH-estimation}. 
	\item We show that the proposed power allocation problem is non-convex in general. Thus, we develop a novel solution method based on iteratively solving a convex optimization instance even when the RF energy conversion model is non-linear. We show that our proposed algorithm can achieve a solution that possesses the desired optimality. We provide a closed-form solution for the cases where we can simplify the EH model to be linear, and the BS has massive antennas and uses orthogonal pilot transmission combined with LS channel estimation and maximum ratio transmission. 
\end{itemize}

\subsection{Paper Structure}
The rest of the paper is organized as follows. 
We describe our WPSN system model and formulate a novel power allocation problem in Section~\ref{sec:Modelling-and-Problem-Formulation}. We study the problem and provide a solution approach and the corresponding analysis of the approach in Section~\ref{sec:Single-BS-Case}, followed by the numerical results in Section~\ref{sec:numerical-results}. We conclude this paper in Section~\ref{sec:conclusions}. To improve the readability of the paper, we present all the proofs in Appendix~\ref{appendix:proofs}. 

In this paper, we use the notations as follows: For a vector $\vec{x}$, $\vec{x}^T$ and $\vec{x}^H$ is its transpose and conjugate transpose, respectively. Notation $\|\vec{x}\|^2\triangleq\vec{x}^H\vec{x}$. Notation $\xrightarrow{\text{a.s.}}$ means converge almost surely.  Table~\ref{table:notation} summarizes the main notations of the paper.

\begin{table}	
	\centering
	\caption{Main notations used throughout the paper.}\label{table:notation}
	\begin{tabular}{|c|l|}
			\hline \textbf{Symbols} & \textbf{Definition} \\ 			
			\hline $E$ & Total energy to be allocated in a time block\\
						\hline $E_{i}^{\rm{t}}$ & Energy to be transmitted to $v_i$ from the BS\\
						\hline $E_{\rm{s}}^*(w)$ & \tabincell{l}{Minimum total energy to satisfy the network sensing rate $w$} \\
			\hline $N$ & Number of sensor nodes \\
			\hline $N_{\rm{t}}$ & Number of antennas of each BS\\
			\hline $P^{\rm{p}}$ & Power of the pilots for channel acquisition by the BS\\
			\hline $c_i$ & 	Static power consumption of $v_i$\\
			\hline $e_{i}$ & Power consumption of $v_i$ to transmit a unit size data\\
			\hline $g_{i}(P^{\rm{p}})$ & Expected pilot-estimation-beamforming gain of the BS to $v_i$\\
			\hline $\vec{h}_{i}$ & Channel from the BS to $v_i$\\
			\hline $w_i$ & Sensing rate of $v_i$\\
			\hline $w_{\min}$ & Minimum of the sensing rates of all nodes, i.e., $w_{\min}=\min_i\{w_i\}$\\
			\hline $\eta(\cdot)$ & Function of the RF-DC conversion of the nodes\\		
			\hline
	\end{tabular}
\end{table}

\section{Modelling and Problem Formulation}\label{sec:Modelling-and-Problem-Formulation}
In this section, we introduce our WPSN setting. We then formulate a novel power allocation problem and analyze its complexity.

\subsection{Network Model}
We consider a WPSN as shown in Fig.~\ref{Fig:NetSys}. The network has one BS and $N>1$ homogeneous sensor nodes, $v_1, v_2,\ldots, v_N$, demanding low energy consumption rates. We denote set $\mathcal{N}=\{1,2,\ldots,N\}$. The BS has $N_{\rm{t}}$ antennas and uses energy beamforming to transmit RF energy to the sensor nodes. Accordingly, the sensor nodes use the harvested energy to sense and to transmit data. Here, we consider a star topology for the network where the sensor nodes transmit their measurements directly to the data sink, and we comment on how our framework easily applies to a general mesh topology in Section~\ref{sec:sub:power_allocation_problem}. We assume that the design of energy transmit waveform and channel estimation have been pre-decided, and they are the input of the problem. Here, we optimize the power allocation of the BS in pilot signals (for channel estimation) and in the energy transmission. 

\subsection{Channel Estimation and Beamforming}\label{sec:CH-estimation}
\begin{figure}[t]
		\centering
		\includegraphics[width=0.45\textwidth]{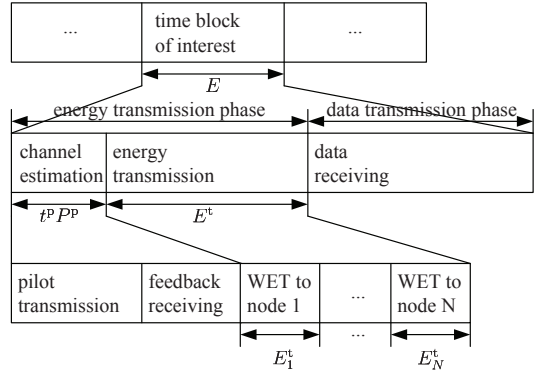}
			\vspace{-0.2cm}
		\caption{In each time block of interest, the base station allocates its power for estimating the channels of the sensor nodes, and the energy to transmit to each sensor node.}
		\label{Fig:powerallocate}  
		\vspace{-0.8cm}
	\end{figure}

We consider a block-fading wireless channel where the channels from the BS to the sensor nodes remain constant during a coherence interval, hereafter called time block~\cite{xu2016general}. The energy is carried by a single-carrier signal. We normalize the length of a time block to be 1. In each time block of interest, the BS first sends pilot signals in the initial $t^{\rm{p}}$ time with power $P^{\rm{p}}$, receives feedbacks from the sensor nodes, and gets the estimation of the channel. Such an approach is similar to the forward-link training and the power probing scheme in~\cite{zeng2017communications}. Our framework allows for both 1) the sensor nodes estimate the channel and feed the estimation back, and 2) the sensor nodes feed the received signal back and the BS estimates the channel. For both cases, the quantization and noise at the feedback process can be considered as additional white noises, which do not affect the results of the paper. In the remaining $1-t^{\rm{p}}$ time, the BS transmits energy $E_{i}^{\rm{t}}$ to each sensor node $i\in\mathcal{N}$, as shown in Fig.~\ref{Fig:powerallocate}.  We assume that the BS uses a time-splitting scheme to charge the sensor nodes. More specifically, within a time block of interest, the BS first transmits energy to $v_1$, then to $v_2$, and so on. Although such a time-splitting scheme is suboptimal, it has lower complexity and the performance is close to the optimum~\cite{du2018towards}. 

 Let  
$E$ be the total energy in the time block for channel estimation and energy transmission. Then, we have $E=t^{\rm{p}}P^{\rm{p}}+\sum_jE_{i}^{\rm{t}}$.  Based on the estimation of the channels, the BS can form energy beams using the existing approaches~\cite{zeng2017communications,du2018towards,alsaba2018beamforming}. We consider the energy transmission as two consecutive processes. The first process denotes how much RF energy is received at the antenna of each node $i$, $E_i^{\rm{r}}$, as a function of how much energy is transmitted from the BS, $E_i^{\rm{t}}$. According to the study in~\cite{clerckx2018beneficial}, we have that $E_i^{\rm{r}}=g_{\textrm{TX-RX},i}E_i^{\rm{t}}$,\footnote{We should note that this model is also valid for modulated single-carrier signals, and also the multisine carrier signals with uniform power allocation on each antenna when the channel is frequency-flat, since the performance of this uniform power allocation is very closes to the optimum in frequency-flat channels~\cite{clerckx2016waveform}.} which is a proportional relation and the gain $g_{\textrm{TX-RX},i}$ depends on the exact channel gain, the accuracy of the channel estimation, and the transmitted RF signals that carry the energy. The output of the first process will be the input of the second process, which denotes how much energy is harvested by node $i$, $E_i^{\rm{h}}$, as a function of $E_i^{\rm{r}}$. We represent it by $E_i^{\rm{h}}=\eta(E_i^{\rm{r}})$, where $\eta(\cdot)$ is the RF-DC conversion function that depends on the rectenna model (here we assume that the recetenna circuits of all sensor nodes are the same). We describe the details of the two processes in the following.

In the first process, we call a combination of the approaches for pilot transmission, channel estimation, and energy beamforming as a pilot-estimation-beamforming (PEB) scheme (here, we do not limit to any particular PEB scheme, but we formulate a general approach that can be applied to different cases of practical interest as long as they satisfy the qualification conditions that will be described in Assumption~\ref{assumpt:g} and~\ref{assumption:eta}). Recall that $g_{\textrm{TX-RX},i}$ depends on the accuracy of the channel estimation, which relates to the power of transmitted pilots $P^{\rm{p}}$. Thus, we model $g_{\textrm{TX-RX},i}=g_i(P^{\rm{p}})$, and call it PEB gain.\footnote{The gain also depends on other factors, such as modulation of the signals and channel estimation methods. However, such factors are considered as input rather than decision variables in our paper. Thus, they are abstracted by the function $g_i(\cdot)$.}  We give an example of $g_i$ for the cases of single continuous wave energy beamforming as follows: 

Consider a block-fading channel $\vec{h}_{i}$ from the BS to $v_i$ with an additive white Gaussian noise. The covariance of the noise is $\sigma_{\rm{n}}^2\vec{I}$. In a PEB scheme, the BS transmits pilots with power $P^{\rm{p}}$. After receiving $N_{\rm{t}}$ pilots, $v_i$ has its received signal and then transmits it back. Based on the feedback, the BS makes an estimation of the channel $\hat{\vec{h}}_{i}(P^{\rm{p}})$, which is a function of $P^{\rm{p}}$. Then, during the energy transmission,  if the BS transmits energy $E_{i}^{\rm{t}}$ to $v_i$ with beamforming ${\vec{b}}_{i}(\hat{\vec{h}}_i(P^{\rm{p}}))$ (we simplify it as ${\vec{b}}_{i}(P^{\rm{p}})$ in the following for the notation convenience),  
the received energy of $v_i$ would be
$E_{i}^{\rm{r}}=E_{i}^{\rm{t}}\|{\vec{b}}_{i}^H(P^{\rm{p}})\vec{h}_{i}\|^2/\|{\vec{b}}_{i}(P^{\rm{p}})\|^2$. Thus, the PEB gain in this example is $g_i(P^{\rm{p}})=\|{\vec{b}}_{i}^H(P^{\rm{p}})\vec{h}_{i}\|^2/\|{\vec{b}}_{i}(P^{\rm{p}})\|^2$.

As we can see, the PEB gain $g_{i}(\cdot)$ abstracts the combined effects of the PEB scheme. Any set of schemes (including the scheme of power probing~\cite{zeng2017communications}) that satisfy the following conditions are compatible with our framework.


\begin{assumption}[Beamforming qualification condition]
\label{assumpt:g}
We assume that for the BS and any sensor node $i$, and any pilot transmission power $P^{\rm{p}} >0$, the PEB gain $g_{i}(P^{\rm{p}})$ satisfies the following conditions: i) $g_{i}(\cdot)$ is an increasing non-negative and bounded function w.r.t. $P^{\rm{p}}$, and ii) it is smooth and concave w.r.t. $P^{\rm{p}}$, thus $g'_{i}(\cdot)>0$, and $g''_{i}(\cdot)\leq 0$.
\end{assumption}
\begin{rem}
The example we provided is for unmodulated continuous wave beamforming. Regarding modulated energy-carrier signals and multisine signals, we are not sure whether the beamforming of these signals would satisfy Assumption~\ref{assumpt:g}, due to the lack of research on how the accuracy of the channel would affect the received energy under such beamformings. In fact, we do not limit the algorithm to a specific channel estimation method or beamforming scheme, in which case the developed algorithms  may not be applicable for other channel estimation methods. Instead, our results are valid for any PEB scheme that satisfies the beamforming qualification condition of Assumption~\ref{assumpt:g}. We provide examples in Section~\ref{sec:LSEstimation} where we show its validity for orthogonal pilot transmission combined with LS channel estimation or mimimum mean square error (MMSE) estimation~\cite{biguesh2006training}, and maximum ratio transmission beamforming~\cite{lo1999maximum}. For any other PEB schemes that pass this condition and the EH condition that will be describe next, one can find the optimal power allocation for channel estimation and beamforming based on the method provided here. For those that do not pass this beamforming qualification condition, we will provide some discussions in Remark~\ref{rem:concavity}.
\end{rem}

From the monotonicity of $g_{i}$, it is straightforward to know that, the PEB gain $g_{i}$ that the BS can have is bounded by a lower bound $0$, and an upper bound $g_i(E/t^{\rm{p}})$. Such a lower bound and upper bound will be used in the development  of our solution algorithm, as described in Section~\ref{sec:Single-BS-Case}.
 

Next, we consider the RF-DC conversion process, and we assume that the EH model $\eta(\cdot)$ satisfies the following condition:
\begin{assumption}[Energy harvesting condition]
\label{assumption:eta}
The mapping from  the received RF energy to the harvested energy, i.e., $E^{\rm{h}}=\eta(E^{\rm{r}})$ is a non-negative and monotone increasing function, w.r.t. the received energy $E^{\rm{r}}$.
\end{assumption}
\begin{rem}
This model says that, for a given input signal/waveform and a realization of the wireless channels, the RF-DC circuit of a node can harvest more RF energy  with more received RF energy at the antenna. Many non-linear RF EH models~\cite{boshkovska2015practical,clerckx2018beneficial,clerckx2018wireless} (for both singlesine or multisine carriers with some given waveform design schemes, e.g. the uniform power, adaptive single sinewave, and adaptive matched filter in~\cite{clerckx2016waveform}, as discussed in Appendix~\ref{appendix:adapative-waveform}) and linear models~\cite{sun2014joint,du2018towards,kashyap2016feasibility} (i.e., $E_i^{\rm{h}}=\eta E_i^{\rm{r}}$) can satisfy this assumption. However, it should be noticed that not all beamforming schemes and RF-DC circuits follow the assumption, e.g., when the transmitter could use adaptive waveform design that suboptimally allocates power to different waveforms according to the channel state.The theoretical results of the paper are valid whenever Assumptions~\ref{assumpt:g} and~\ref{assumption:eta} are satisfied. When they are not satisfied, the solution approach may not work and it will be an open question for these cases. In addition, when the transmit waveform strategy of the BS is not predefined and static, the assumption might not hold if there exists a candidate transmit waveform strategy in the strategy set that violates the assumption.
\end{rem}

\subsection{Energy Consumption Model}
Each sensor node $i$ uses a predefined fixed power to transmit data in a predetermined data rate. We denote the energy consumption to sense, to process, and to transmit a unit data to the sink node by $e_i>0$. Besides, its static energy consumption is $c_i$, which accounts for circuits consumptions and also the power of sending channel feedback to the BS.  Denote the sensing rate of node $v_i$ by $w_i$. Then, we have that the total energy consumption of $v_i$ is $e_iw_i+c_i$.\footnote{This model is widely used for WSNs~\cite{Chang04,xie2015mobile,koutsopoulos2014distributed} because the power that can be used for data transmission by the sensor nodes is very limited, compared to other wireless devices such as mobile phones. However, if one uses the model based on Shannon capacity, our approach is still valid with proper modifications.} We require that the average consumed energy of each node is no larger than the average harvested energy, i.e., $e_iw_i+c_i\leq E_{i}^{\rm{h}},\forall i$. Under this requirement, we will optimize the monitoring performance of the WPSN, as we describe next.

\subsection{Power Allocation Problem}
\label{sec:sub:power_allocation_problem}
The monitoring performance of the WPSN considered here depends on how many measurements are received at the sink node. Naturally, we hope that the nodes make as many measurements as possible. Besides, we do not want to have some nodes to make little measurements whilst some other nodes make too many measurements. Thus, we use the minimum of the sensing rate of all sensor nodes, $w_{\min}\triangleq\min_i\{w_i\}$, as the monitoring performance metric of the WPSN. We also call it network sensing rate. Denote $\vec{w}=[w_1,w_2,\ldots,w_N]^T$ and $\vec{E}^{\rm{t}}=[E_{1}^{\rm{t}},\ldots, E_{N}^{\rm{t}}]^T$. Then, we are ready to formulate the power allocation problem as follows:
 \begin{subequations}
 	\label{Problem0}
 	\begin{alignat}{3}
 	\max_{w_{\min},\vec{w},\vec{E}^{\rm{t}}, P^{\rm{p}}} &\quad  && w_{\min}\\
 	\text{s.t.}& && w_i\geq w_{\min}\,,\forall i\in\mathcal{N},\label{Problem0:wi}\\
 	& && e_iw_i+c_i\leq \eta\left(E_{i}^{\rm{t}}g_{i}(P^{\rm{p}})\right)\,,\forall i\in\mathcal{N},\label{Problem0:energy}\\
 	& && t^{\rm{p}}P^{\rm{p}}+\sum_iE_{i}^{\rm{t}}\leq E\,,\label{Problem0:Psum}\\
 	& && w_i,E_{i}^{\rm{t}},P^{\rm{p}}\geq 0\,,\forall i\in\mathcal{N}\,,\label{Problem0:positive}
 	\end{alignat}
 \end{subequations}
 where the objective is to maximize the network sensing rate, $w_{\min}$; Constraint~\eqref{Problem0:energy} is the energy causality, i.e., the consumed energy of a node must be no larger than the energy it harvests; Constraint~\eqref{Problem0:Psum} is the power limit of the BS; and Constraint~\eqref{Problem0:positive} is the non-negative constraint of the decision variables. The problem is to allocate the power of channel estimation and the energy to be transmitted to each sensor node, such that the network sensing rate is maximized. 

\begin{rem}
The optimal solution of Problem~\eqref{Problem0} should hold that $w_i=w_{\min},\forall i\in\mathcal{N}$. This can proved by contradiction. Briefly speaking, if there is a node with a sensing rate larger than $w_{\min}$, the BS can transmit less energy to this node, and increase the transmission energy to other nodes to increase their sensing rates, and thus the network sensing rate.
\end{rem}
\begin{rem}[Extension to mesh networks]
For a mesh sensor network (i.e., the sensor nodes will relay other nodes measurements) with a fixed routing, we only need to modify parameter $e_i$ as the summation of the power consumption of sensing, processing, and transmitting its measurement to its destination, and the power consumption of receiving and relaying measurement from each of its child node to its destination. Then, the results of the paper still hold.
\end{rem} 

\subsection{Complexity Analysis}
From Constraint~\eqref{Problem0:energy}, even when the RF-DC conversion function $\eta(\cdot)$ is linear, simple algebra shows that  the Hessian matrix
is not necessarily positive semidefinite. Consequently, Problem~\eqref{Problem0} is not a convex optimization and the solution approach is non-trivial. Notice also that when $N=1$, such a problem can be simplified to a convex optimization. Therefore, the difficulty of Problem~\eqref{Problem0} mainly comes from the power allocation for multiple sensor nodes, addressing which is one of the major technical novelty of this paper. In addition, the non-linear behaviour of $\eta(\cdot)$ makes the problem even more challenging. Despite the non-convexity of the resulting optimization problem, we propose an efficient algorithm that finds the optimal solution of Problem~\eqref{Problem0}, as will be presented in the next section.

\section{Solution Method}\label{sec:Single-BS-Case}
In this section, we investigate a solution approach to solve Problem~\eqref{Problem0}. Then, we show the convergence properties and the computational complexity of the algorithm. We then provide some illustrative examples, in which we can find closed-form solutions of~\eqref{Problem0}. Last, we will discuss some special cases, such as the problem with linear EH models.
\subsection{Algorithm Development}
%
To develop the solution algorithm, we first study a sub-problem on checking the feasibility of the total energy $E$, based on the assumption that $w_{\min}$ is given. Assume that $w_{\min}$ is given. The following sub-problem finds the minimum total energy to satisfy $w_{\min}$:
\begin{subequations}
\label{ProblemSub_NonLinear}
\begin{alignat}{3}
\min_{E_{\rm{s}},\vec{E}^{\rm{t}},P^p} &\quad && E_{\rm{s}}\\
\text{s.t.}&  &&  e_iw_{\min}+c_i\leq \eta \left( E_i^{\rm{t}}g_i(P^{\rm{p}})\right),\forall i\in\mathcal{N}\,,\label{ProblemSub_NonLinear:EnergyCasuality}\\
& && t^{\rm{p}}P^{\rm{p}}+\sum_{i\in\mathcal{N}}E_i^{\rm{t}}\leq E_{\rm{s}}\,,\\
& && P^{\rm{p}},E_i^{\rm{t}}\geq 0,\forall i\in\mathcal{N}\,.
\end{alignat}
\end{subequations}

Since $\eta(\cdot)$ is non-negative and monotone increasing, we have that $\eta(\cdot)$ has an inverse function, denoted by $\eta^{-1}(\cdot)$, and $\eta^{-1}(\cdot)$ is also non-negative and monotone increasing. Therefore, Constraint~\eqref{ProblemSub_NonLinear:EnergyCasuality} gives us that $E_i^{\rm{t}}\geq \eta^{-1}(e_iw_{\min}+c_i)/g_i(P^{\rm{p}})$. To make it more concise, we define $f_i(P^{\rm{p}};w_{\min})\triangleq\eta^{-1}(e_iw_{\min}+c_i)/(g_i(P^{\rm{p}}))$. Then, Problem~\eqref{ProblemSub_NonLinear} is equivalent to the following one:
	\begin{align}	\label{ProblemSub1_Nonlinear}
	\min_{0\leq P^{\rm{p}}\leq E/t^{\rm{p}}}\quad  & E_{\rm{s}}(P^{\rm{p}};w_{\min})\triangleq t^{\rm{p}}P^{\rm{p}}{+}\sum_{i\in\mathcal{N}}f_i(P^{\rm{p}};w_{\min})\,,
	\end{align}
where $E_{\rm{s}}(P^{\rm{p}};w_{\min})$ is the total energy to satisfy the required sensing rate $w_{\min}$, when the BS uses power $P^{\rm{p}}$ for the channel estimation. We denote by $E^*_{\rm{s}}(w_{\min})$ the optimum of Problem~\eqref{ProblemSub1_Nonlinear}, then we have the following proposition for Problem~\eqref{ProblemSub1_Nonlinear}, proved in Appendix:


\begin{prop}
\label{prop:convex}
Problem~\eqref{ProblemSub1_Nonlinear} is a single variable convex optimization problem, and the optimal solution is either at $P^{\rm{p}}=0$ or at the point where its derivative $E_{\rm{s}}'(P^{\rm{p}};w_{\min})$ is 0. 
\end{prop}
\begin{rem}
\label{rem:E's}
(The proof of) Proposition~\ref{prop:convex} implies that $E'_{\rm{s}}(P^{\rm{p}};w_{\min})$ is monotone increasing with $P^{\rm{p}}$. Thus, we have that
\begin{itemize}
\item  If $E'_{\rm{s}}(0;w_{\min})=t^{\rm{p}}+\sum_{i\in\mathcal{N}}f'_i(0;w_{\min})\geq 0$ for all $P^{\rm{p}}\geq 0$, then the optimal solution of Problem~\eqref{ProblemSub1_Nonlinear} is $P^{\rm{p}}=0$.
\item If $E'_{\rm{s}}(E/t^{\rm{p}};w_{\min})\leq 0$, then due to the monotonicity of $E'_{\rm{s}}(P^{\rm{p}};w_{\min})$, we have that the $P^{\rm{p}}$ that satisfies Equation~\eqref{Eqn:OptPp} is larger than $E/t^{\rm{p}}$. This means that the given $w_{\min}$ is not achievable with the total energy constraint $E$.
\item Otherwise, i.e., $E'_s(0;w_{\min})<0<E'_s(E/t^{\rm{p}};w_{\min})$, we can achieve the unique solution of Equation~\eqref{Eqn:OptPp} numerically using a bisection search algorithm in the region $P^{\rm{p}}\in[0,E/t^{\rm{p}}]$.
\end{itemize} 
\end{rem}

Let $E_{\rm{s}}^*(w)$ be the optimum of Problem~\eqref{ProblemSub1_Nonlinear} given $w$, and $w^*_{\min}$ be the optimum of Problem~\eqref{Problem0}. If $E_{\rm{s}}^*(w)<E$, we have that $w<w^*_{\min}$; otherwise $w\geq w^*_{\min}$. This gives us the solution algorithm for Problem~\eqref{Problem0} based on bisection searching. The idea is as follows:

We first find a lower bound and an upper bound of $w_{\min}$, which is denoted by $w_{\min}^{\rm{l}}$ and $w_{\min}^{\rm{u}}$ respectively. For the lower bound, we can easily choose $w_{\min}^{\rm{l}}=0$. For the upper bound, it corresponds to the case that the BS can achieve the upper bound of the PEB gain, i.e., $g_i(E/t^{\rm{p}})$, without spending any power in pilot transmission. Thus, one can choose $w_{\min}^{\rm{u}}$ be the optimal solution of the following linear optimization problem:
\begin{subequations}
	\label{ProblemUpperbound_NonLinear}
	\begin{alignat}{3}
	\max_{w_{\min},\vec{E}^{\rm{t}}} &\quad&& w_{\min}\\
	\text{s.t.}& && e_iw_{\min}+c_i\leq\eta_{\max} E_i^{\rm{t}} g_{i}\left(\frac{E}{t^{\rm{p}}}\right),\forall i\in\mathcal{N}\,,\label{ProblemUpperbound_NonLinear:C1}\\
	& && \sum_{i\in\mathcal{N}} E_i^{\rm{t}}\leq E\,,\label{ProblemUpperbound_NonLinear:C2}\\
	& && w_{\min},P_i^{\rm{t}}\geq 0,\forall i\in\mathcal{N}\,,
	\end{alignat}
\end{subequations}
where $\eta_{\max}\triangleq \max_{x\geq 0} \eta(x)/x$, such that $\eta(x)\leq \eta_{\max}x$ for $x\geq 0$.
This gives us the  proposition of the upper bound rate as follows:

\begin{lemma}
\label{prop:wupperbound}
Consider a feasible Problem~\eqref{Problem0}. The upper bound of $w_{\min}$ is given by
\begin{align}
\label{Eqn:wupperbound}
w_{\min}^{\rm{u}}=\frac{\eta_{\max} E-\sum_{i\in\mathcal{N}}\frac{c_i}{g_{i}(\frac{E}{t^{\rm{p}}})}}{\sum_{i\in\mathcal{N}}\frac{e_i}{g_{i}(\frac{E}{t^{\rm{p}}})}}\,. 
\end{align}
\end{lemma}

Once we have known the upper bound and the lower bound of $w_{\min}$, we can check the feasibility of $w_{\min}=0.5(w_{\min}^{\rm{l}}+w_{\min}^{\rm{u}})$ for Problem~\eqref{ProblemSub1_Nonlinear}. If $w_{\min}$ is feasible, we update the new lower bound by $w_{\min}$; otherwise, we update the new upper bound by $w_{\min}$. This proceeds iteratively until the lower bound and the upper bound converge. Algorithm~\ref{Alg:PowerAllocation} summarizes this procedure.

\begin{rem}
\label{rem:concavity}
For the cases where the PEB gain $g_i(P^{\rm{p}})$ is not concave w.r.t. the pilot power, it is not sufficient to achieve that Problem~\eqref{ProblemSub1_Nonlinear} is convex. However, since Problem~\eqref{ProblemSub1_Nonlinear} is a single variable optimization in a bounded region with no other constraints, one can use numerical approaches, such as bisection search and Newton's method, to find a solution that is close to the optimum. Thus, we can still use Algorithm~\ref{Alg:PowerAllocation} to find the solution of Problem~\eqref{Problem0} by properly modifying its Line 11.
\end{rem}

	\begin{algorithm}[t]
		\caption{Solution for Problem~\eqref{Problem0}}\label{Alg:PowerAllocation}
		\linespread{1}\selectfont
		\begin{algorithmic}[1]\small
			\REQUIRE $e_i,c_i, g_i(\cdot),  \forall i\in\mathcal{N}$, $E, t^{\rm{p}}, \varepsilon, \eta(\cdot)$
			\ENSURE  $E_i^{\rm{t}},\forall i\in\mathcal{N}, P^{\rm{p}},w_{\min}$

\STATE Set $w_{\min}^{\rm{l}}=0$
			\IF {$\sum_{i\in\mathcal{N}} c_i/( g_{i}(E/t^{\rm{p}}))>\eta_{\max}E$}
			\STATE The problem is infeasible and return $w=0$.
			\ELSE
			\STATE Find $w_{\min}^{\rm{u}}$ according to Equation~\eqref{Eqn:wupperbound}
			\WHILE {$w_{\min}^{\rm{u}}-w_{\min}^{\rm{l}}\geq \varepsilon$}
			\STATE Set $w_{\min}=0.5(w_{\min}^{\rm{u}}+w_{\min}^{\rm{l}})$
			\IF {$t^{\rm{p}}+\sum_{i\in\mathcal{N}} f_i'(0;w_{\min})\geq 0$}
			\STATE $P^{\rm{p}}\leftarrow 0$, $E_i^{\rm{t}}\leftarrow f_i(0;w_{\min}),\forall i\in\mathcal{N}$, $E_{\rm{s}}^*(w_{\min})=\sum_{i\in\mathcal{N}}E_i^{\rm{t}}$
			\ELSE
			\STATE Find $P^{\rm{p}}$ that satisfies Equation~\eqref{Eqn:OptPp}. $E_i^{\rm{t}}\leftarrow f_i(P^{\rm{p}};w_{\min}),\forall i$, $E_{\rm{s}}^*(w_{\min})=t^{\rm{p}}P^{\rm{p}}+\sum_{i\in\mathcal{N}}E_i^{\rm{t}}$
			\ENDIF
			\IF {$E_{\rm{s}}^*(w_{\min})-E>0$}
			\STATE Update $w_{\min}^{\rm{u}}=w_{\min}$
			\ELSE
			\STATE Update $w_{\min}^{\rm{l}}=w_{\min}$			
			\ENDIF
			\ENDWHILE
			\STATE Set $w_{\min}=w_{\min}^{\rm{l}}$, and set $P^{\rm{p}}$ as the optimal solution of convex optimization Problem~\eqref{ProblemSub1_Nonlinear}
			\FOR {$i=1$ to $N$}		
			\STATE Set $E_i^{\rm{t}}=f_i(P^{\rm{p}};w_{\min})$.
			\ENDFOR
			\RETURN $E_i^{\rm{t}},P^{\rm{p}},\forall i\in\mathcal{N}, w_{\min}$.
			\ENDIF
		\end{algorithmic}
	\end{algorithm}

\subsection{Performance Analysis}
Now, we are ready to analyze the performance of Algorithm~\ref{Alg:PowerAllocation}  in solving Problem~\eqref{Problem0}, in terms of the optimality of the final solution and its computational complexity. 
The near optimality of the algorithm is given by the following proposition:
\begin{prop}
\label{thm:Th1}
Let Problem~\eqref{Problem0} be feasible and let its optimum be $w_{\min}^{\rm{o}}>0$. Given any arbitrary small gap $\varepsilon$, Algorithm~\ref{Alg:PowerAllocation} finds a feasible solution $(w_{\min},\vec{w},\vec{E}^{\rm{t}},P^{\rm{p}})$ that satisfies $w_{\min}^{\rm{o}}-w_{\min}< \varepsilon$.
\end{prop}

%
%

Regarding the complexity of Algorithm~\ref{Alg:PowerAllocation}, we have the following proposition:
\begin{prop}\label{prop:time-complexity}
Let Problem~\eqref{Problem0} be feasible and let its optimum be $w>0$. Given the arbitrary optimality gap $\varepsilon$, the time complexity of Algorithm~\ref{Alg:PowerAllocation} is at most $\mathcal{O}\left(N\log(E)\log(E/(N\varepsilon))\right)$, where recall that $N$ is the number of sensor nodes, and $E$ is the total energy of BS for each time block.
\end{prop}

%

Consequently, we conclude that Algorithm~\ref{Alg:PowerAllocation} is an efficient approach (sublinear in $N$) to find a feasible and near optimal solution  (arbitrarily close to the optimal $w_{\min}$). However, for the cases where Assumptions~\ref{assumpt:g} and~\ref{assumption:eta} are not satisfied, the optimality gap and time complexity of Algorithm~\ref{Alg:PowerAllocation} will depend on the the objective function and the solution approach of Problem~\eqref{ProblemSub1_Nonlinear}. For example, if the objective function of Problem~\eqref{ProblemSub1_Nonlinear} is unimodal, then we can still use bisection method  in Line 11 of Algorithm~\ref{Alg:PowerAllocation} and observe the same time complexity as in Proposition~\ref{prop:time-complexity}. Next, we provide an illustrative example where  the BS uses a simple LS algorithm to estimate the CSI  and then uses maximum ratio beamforming for the energy transmission.

\subsection{Illustrative Example}
\label{sec:LSEstimation}
In this subsection, we will give a simple example that the BS uses orthogonal pilot transmission combined with LS channel estimation and maximum ratio transmission, to show this PEB scheme follows qualification conditions of Assumption~\ref{assumpt:g}. In this special case, we can further simplify the solutions of our iterative optimization approach and derive closed-form expressions.  

\begin{figure}[t]
		\centering
		\subfigure[]{
				\includegraphics[width=0.43\textwidth]{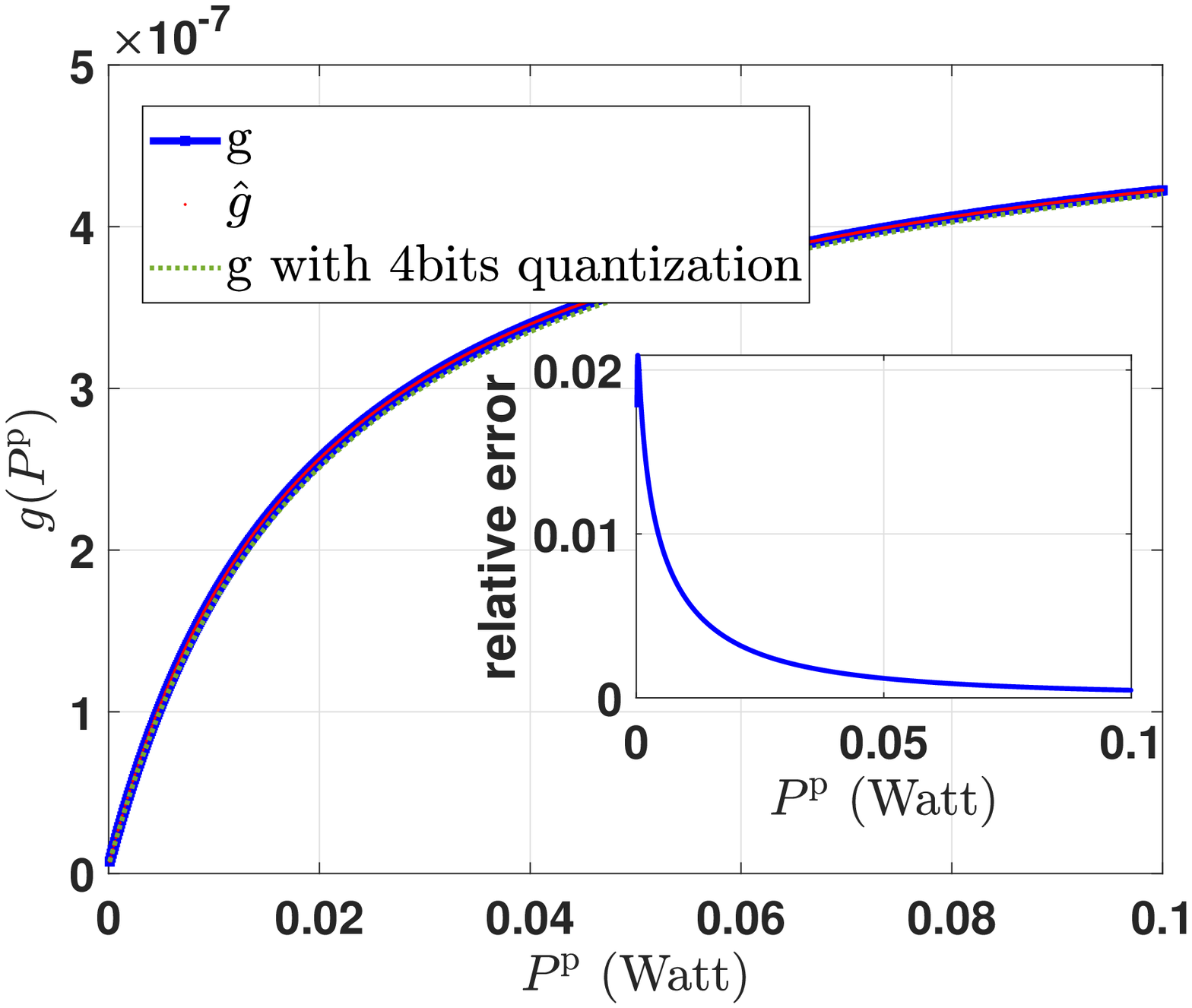}\label{Fig:G_LS}}
			\subfigure[]{\includegraphics[width=0.43\textwidth]{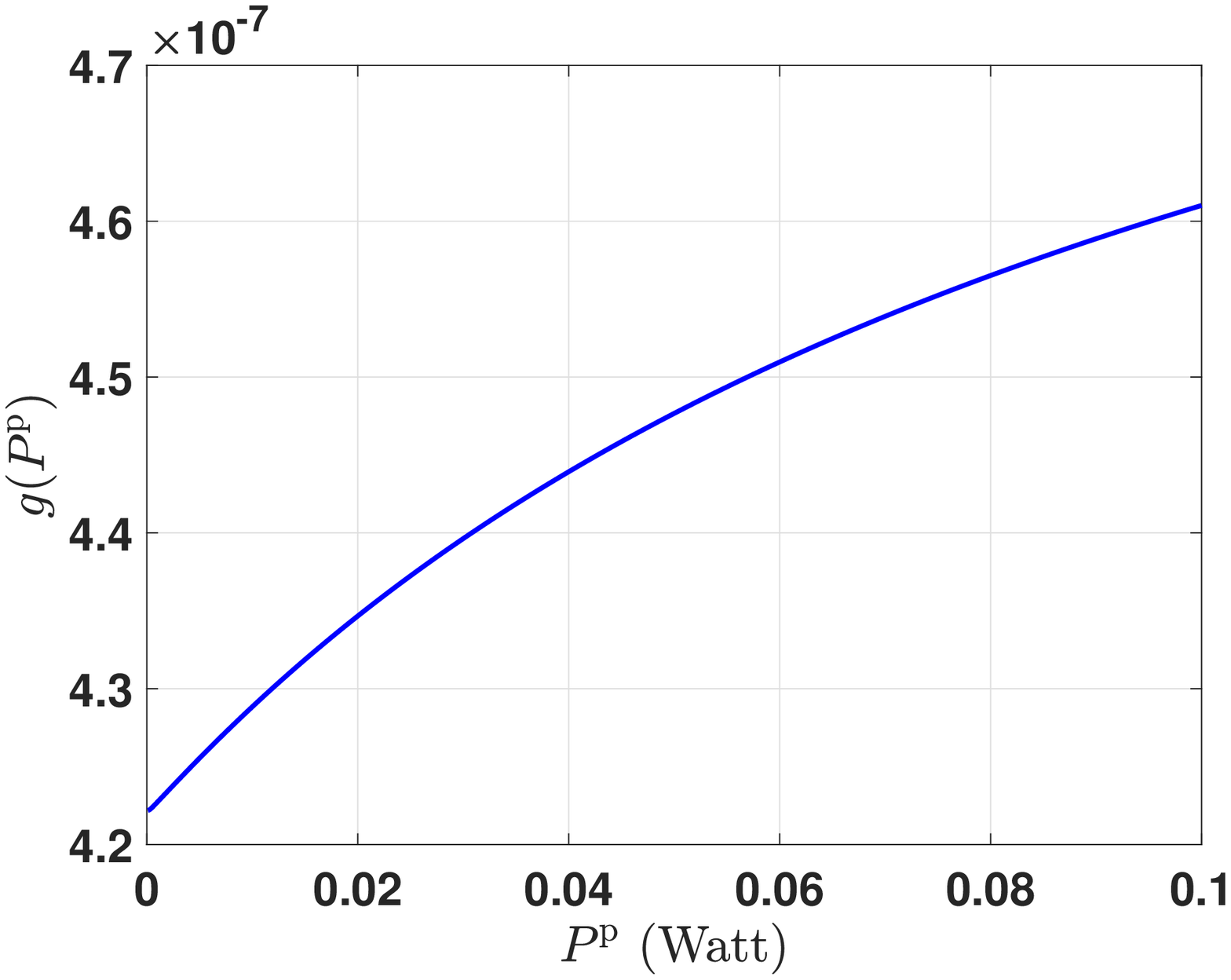}\label{Fig:G_MMSE}}
		\caption{(a) The PEB gain and its approximation at different pilot power $P^{\rm{p}}$ of the least square estimator cases; (b) The PEB gain at different pilot power $P^{\rm{p}}$ of the MMSE estimator cases. }
		\vspace{-0.8cm}
	\end{figure}

First, let us derive the expected harvested energy when the BS uses the LS estimation. During the channel acquisition phase, the sensor nodes use a switch circuit to connect their antenna element with their communication module~\cite{zeng2017communications}. Recall that the BS has $N_{\rm{t}}$ antennas, and each sensor node has one antenna. Thus, the channel from the BS to node $i$ is $\vec{h}_i$ of size $N_{\rm{t}}\times 1$, and we assume that $\vec{h}_i$ is independent to $\vec{h}_k$ for $i\neq k$. To estimate the channels toward all the nodes, the BS broadcasts $N_{\rm{t}}$ pilots to the nodes. For simplicity, we assume that the pilots are the column vectors of the identical matrix $\vec{I}_{N_{\rm{t}}}$. If the power of the pilots are $P^{\rm{p}}$, then for each node $i$, it receives the signal as $\vec{y}_i=\sqrt{P^{\rm{p}}/N_{\rm{t}}}\vec{I}\vec{h}_i+\vec{n}_i$, where $\vec{n}_i$ is an additive white Gaussian noise at the node $i$ with covariance $\sigma_{\rm{n}}^2\vec{I}$. Consider the feedback with quantization, then the feedback of $\vec{y}$ is $\hat{\vec{y}}_i=\vec{y}+\vec{e}_{\textrm{q},i}$, where $\vec{e}_{\textrm{q},i}$ is the zero-mean quantization error. We assume that the quantization is i.i.d, and is independent of the channel and the noise at the receiver. The LS estimation of $\vec{h}$, based on $\hat{\vec{y}}_i$ is 
\begin{align*}
\vec{\hat{h}}_i^{\textrm{LS}} =\left(\frac{P^{\rm{p}}}{N_{\rm{t}}}\vec{I}^{H}\vec{I}\right)^{-1}\sqrt{\frac{P^{\rm{p}}}{N_{\rm{t}}}}\vec{I} \hat{\vec{y}}_i=\sqrt{\frac{N_{\rm{t}}}{P^{\rm{p}}}}\hat{\vec{y}}_i=\vec{h}_i+\sqrt{\frac{N_{\rm{t}}}{P^{\rm{p}}}}(\vec{n}_i+\vec{e}_{\textrm{q},i})\,.
\end{align*}
Define $\tilde{\vec{n}}_i=\vec{n}_i+\vec{e}_{\textrm{q},i}$, and the covariance of $\tilde{n}_i$ be $\sigma_{\tilde{\rm{n}}}^2\vec{I}$. By setting $\vec{b}_{i}=\vec{\hat{h}}_i^{\rm{LS}}$ as the beamforming vector, the expected received power of node $i$ becomes
\begin{align*}
E_i^{\rm{r}}(P^{\rm{p}})=E_i^{\rm {t}}\mathbb{E}\left[\frac{\|\vec{h}_i^H\vec{h}_i+\sqrt{\frac{N_{\rm{t}}}{P^{\rm{p}}}}\vec{h}_i^H\tilde{\vec{n}}_i\|^2}{\|\vec{h}_i+\sqrt{\frac{N_{\rm{t}}}{P^{\rm{p}}}}\tilde{\vec{n}}_i\|^2}\right]\,.
\end{align*}
Therefore, we have that
\begin{align}
\label{Eqn:g_j}
g_i(P^{\rm{p}})=\mathbb{E}\left[\frac{\|\sqrt{P^{\rm{p}}}\vec{h}_i^H\vec{h}_i+\sqrt{N_{\rm{t}}}\vec{h}_i^H\tilde{\vec{n}}_i\|^2}{\|\sqrt{P^{\rm{p}}}\vec{h}_i+\sqrt{N_{\rm{t}}}\tilde{\vec{n}}_i\|^2}\right]\,.
\end{align}
where the expectation is taken over the distribution of $\tilde{n}_i$. This is because, when the BS models the function $g_i(\cdot)$ and formulates the optimization problem, it has not sent out the pilots and the noise is not realized. We show in Appendix~\ref{appendix:concavity} that $g_i(P^{\rm{p}})$ is concave when $P^{\rm{p}}\geq (2\sqrt{3}-1)(N_{\rm{t}}^2\sigma_{\rm{n}}^2)/\|\vec{h}_i\|^2$. For the cases when $P^{\rm{p}}$ is smaller than such a threshold, it is not sufficient to determine whether it is concave or not. To have a better understanding of its concavity, we run a simulation scenario as follows. We place the BS at $(0,0)$, and we put a sensor node at a random location near the BS. The distance of the node to the BS turns out to be 11.69~meters. The BS has $N_{\rm{t}}=100$ antennas, transmitting energy pilot at frequency 915MHz. The wireless channel is modelled as Rician with factor 10, and we generate 1000 instances of the channels $\vec{h}_i$. The noise at the sensor node is $\sigma_{\rm{n}}^2=-90$~dBm. We first consider the perfect feedback case, i.e., no quantization error. Then, we vary $P^{\rm{p}}$ from $0.1~mW$ to $0.1W$, and calculate the averaged $g_i(P^{\rm{p}})$ for each $P^{\rm{p}}$. The result is shown in Fig~\ref{Fig:G_LS}. It can be observed that the curve of $g(P^{\rm{p}})$ is an increasing and concave function of $P^{\rm{p}}$.

To simplify the simulations, we want to find an approximation of $g_i(\cdot)$ in~\eqref{Eqn:g_j}. Therefore, we approximate $g_i(P^{\rm{p}})$ by
\begin{align}
g_i(P^{\rm{p}})\approx\hat{g}_i(P^{\rm{p}})=\sigma^2_{h_i}\frac{P^{\rm{p}}\sigma^2_{h_i}+N_{\rm{t}}\sigma^2_{\tilde{\rm{n}}}}{P^{\rm{p}}\sigma^2_{h_i}+N_{\rm{t}}^2\sigma_{\tilde{\rm{n}}}^2}\,,
\end{align}
where $\sigma^2_{h_i}=\mathbb{E}[{\vec{h}_i^H\vec{h}_i}]$.

To compare $g_i(P^{\rm{p}})$ and its approximation $\hat{g}_i(P^{\rm{p}})$, we plot $\hat{g}_i(P^{\rm{p}})$ in Fig~\ref{Fig:G_LS} by the red dots. It can be observed that the two functions are close, i.e., the difference $|\hat{g}(P^{\rm{p}})-g(P^{\rm{p}})|/g(P^{\rm{p}})$ is small. Thus, we consider $\hat{g}_i(P^{\rm{p}})$ as a good approximation of $g_i(P^{\rm{p}})$, and we will use $\hat{g}_i$ instead of $g_i$ in the following and in the simulations. One can examine that, $\hat{g}_i(P^{\rm{p}})$ is a concave and monotone increasing function by simple algebra. We also check the effect of feedback quantization in Fig~\ref{Fig:G_LS}. The green dotted line and the blue solid line in Fig.~\ref{Fig:G_LS} correspond to quantization with 4-bit mantissa and no quantization (perfect feedback), respectively. Observing up to 4\% relative difference between these curves indicates the marginal  effect of quantization compare to that of pilot power optimization. Thus, in the following, we would neglect the quantization error, i.e., $\tilde{\vec{n}}=\vec{n}$,  for the sake of simplicity.

We start from Remark~\ref{rem:E's} and compute $E'(P^{\rm{p}};w_{\min})$. We replace function $g_i$ in Equation~\eqref{Eqn:OptPp} (in Appendix A) by $\hat{g}_i$, and have that
\begin{align}
\label{Eqn:OptPp_LS}
E'(P^{\rm{p}};w_{\min})=t^{\rm{p}}+\sum_{i\in\mathcal{N}}f_i'(P^{\rm{p}}|w_{\min})=t^{\rm{p}}-\sum_{i\in\mathcal{N}}\eta^{-1}\left(e_iw_{\min}+c_i\right) \frac{N_{\rm{t}}\sigma_{\rm{n}}^2(N_{\rm{t}}-1)}{(P^{\rm{p}}\sigma^2_{h_i}+N_{\rm{t}}\sigma_{\rm{n}}^2)^2}\,.
\end{align}
We can see from Equation~\eqref{Eqn:OptPp_LS} that $E'(P^{\rm{p}};w_{\min})$ is strictly increasing with $P^{\rm{p}}$. Therefore, if Equation~\eqref{Eqn:OptPp_LS} is feasible, the $P^{\rm{p}}$ that satisfies $E'(P^{\rm{p}};w_{\min})=0$ is unique, which can be achieved by the bisection approach. For a special case where $\sigma^2_{h_i}=\sigma^2_{h_k}=\sigma^2_{\rm{h}},\forall k\neq i$, (i.e., all the nodes have the same path-loss to the BS), the solution is
\begin{align*}
P^{\rm{p}}=\sqrt{\sum_{i\in\mathcal{N}}\frac{\eta^{-1}(e_iw_{\min}+c_i)N_{\rm{t}}\sigma^2_{\rm{n}}(N_{\rm{t}}-1)}{ t^{\rm{p}}\sigma^4_{\rm{h}}}}-\frac{N_{\rm{t}}\sigma^2_{\rm{n}}}{\sigma^2_{\rm{h}}}\,.
\end{align*}
If we use such a result in Line 11 of Algorithm~\ref{Alg:PowerAllocation}, its computational complexity is then reduced from $\mathcal{O}\left(N\log(E)\log(E/(N\varepsilon))\right)$ to $\mathcal{O}\left(N\log(E/(N\varepsilon))\right)$. We can see that, given a certain channel estimation approach, and some other special conditions, we can revise Algorithm~\ref{Alg:PowerAllocation} to a simpler and faster one. Also, since $\sigma^2_{h_i}=\sigma^2_{h_k}=\sigma^2_{\rm{h}},\forall k\neq i$, we have that $g_i(x)=g_k(x)$. Thus, under such a special case, the upper bound of $w_{\min}$ corresponding to Lemma~\ref{prop:wupperbound} can be simplified to
\begin{align}
\label{Eqn:wmin_Upperbound_LS}
w_{\min}^{\rm{u}}=\frac{\eta_{\max} g\left(\frac{E}{t^{\rm{p}}}\right)P-\sum_{i\in\mathcal{N}} c_i}{\sum_{i\in\mathcal{N}} e_i}\,.
\end{align}
We will use this result in the simulation in Section~\ref{sec:numerical-results} to check the performance of Algorithm~\ref{Alg:PowerAllocation} in the special case of this section.

Besides the LS estimator, we also study the cases where the BS uses the minimum mean square error (MMSE)~\cite{biguesh2006training,bogale2015hybrid} to estimate the channels, i.e., the estimated channel is 
\begin{align*}
\vec{\hat{h}}_{i}^{\rm{MMSE}}=\vec{R}_{h_i}\sqrt{\frac{P^{\rm{p}}}{N_{\rm{t}}}}\left(\frac{P^{\rm{p}}}{N_{\rm{t}}}\vec{R}_{h_i}+\sigma^2_{\rm{n}}\vec{I}\right)^{-1}\vec{y}_i\,,
\end{align*}
where $\vec{R}_{h_i}$ is the channel covariance matrix. Then, similar to what we have done for the LS estimator case, we run the same simulation by setting $\vec{b}_i=\vec{\hat{h}}_{i}^{\rm{MMSE}}$. We present the result in Fig.~\ref{Fig:G_MMSE}, which shows the concavity of the PEB gain $g_i(P^{\rm{p}})$.


\subsection{Solution for Linear Energy Harvesting Model}
Previously, we have provided the solution algorithm for Problem~\eqref{Problem0}. In this subsection, we consider a special case where the EH of the nodes are modelled by a linear function. That is, the EH model is $\eta(x)=\alpha x$, where $\alpha$ is a constant representing the EH rate. In this case, 
we can still use Algorithm~\ref{Alg:PowerAllocation} to solve the problem. The analytical results that we have provided in the previous subsections are also valid for the case of linear EH modelling. It means that, given a non-linear EH model, we can find a linear approximation of the model and compare the solutions of Algorithms~\ref{Alg:PowerAllocation} for the non-linear EH model case with its linear approximation case. Such comparison enables the sensitivity analysis of the solution to the linear approximation.

\subsection{Asymptotic Case}
\label{sec:sub:asymptotic}
In this subsection, we consider the problem with linear EH model we discussed in the previous subsection. We study the behaviour of LS channel estimation and maximum ratio transmit beamforming with $N_{\rm{t}}\rightarrow +\infty$. We have the following results:
\begin{rem}
\label{rem:asymptotic_g}
Consider a BS with sufficiently large number of antennas, constant transmit power per antenna, orthogonal pilot transmission combined with least-square channel estimation and maximum ratio transmission scheme. Then
\begin{align}
\label{Eqn:asymptotic_g}
g_i(P^{\rm{p}})\xrightarrow{\text{a.s.}}\frac{N_{\rm{t}}\sigma_i^4P^{\rm{p}}}{\sigma_i^2P^{\rm{p}}+N_{\rm{t}}\sigma^2_{\rm{n}}}\,,
\end{align}
where $\sigma^2_i=\mathbb{E}_k[\|h_{ik}\|^2]$, and $h_{ik}$ is the $k$-th element of vector $\vec{h}_i$. That is, $g_i(P^{\rm{p}})$ converges almost surely to a monotone increasing and concave function.
\end{rem}

Notice that the total pilot transmit power can grow arbitrarily large in Remark~\ref{rem:asymptotic_g}. We have added that assumption for the sake of mathematical tractability in the proof. In many wireless systems, however, we can replace $N_t$ by the number of multi-path components of the channel, which is a constant number, independent of $N_t$~\cite{moghadam2017pilot}. This number could be very small, e.g., less than 5 in millimeter-wave networks~\cite{rangan2014millimeter}. 

From Remark~\ref{rem:asymptotic_g}, when $N_{\rm{t}}$ is fixed and large enough (e.g., massive antenna regime), the PEB gain $g_i(P^{\rm{p}})$ is monotone increasing and concave with $P^{\rm{p}}$, and therefore it satisfies the qualification conditions of Assumption~\ref{assumpt:g}. Thus, our proposed algorithm is applicable  to such a case. However, in the asymptotic regime, described by Equation~\eqref{Eqn:asymptotic_g}, we can have a simpler solution. Recall that in Algorithm 1, we need to solve sub-problem~\eqref{ProblemSub1_Nonlinear} in each iteration. By substituting~\eqref{Eqn:asymptotic_g} into~\eqref{ProblemSub1_Nonlinear}, the sub-problem turns to be
\begin{align*}
\min_{P^{\rm{p}}\geq 0} \quad t^{\rm{p}}P^{\rm{p}}+\sum_{i\in\mathcal{N}}\frac{\sigma^2_{\rm{n}}(e_iw_{\min}+c_i)}{\alpha\sigma^4_{i}P^{\rm{p}}}+\sum_{i\in\mathcal{N}}\frac{e_iw_{\min}+c_i}{\alpha N_{\rm{t}}\sigma^2_{i}}\,.
\end{align*}
The optimal solution can be easily achieved as $P^{\rm{p}}=\sqrt{\sum_{i\in\mathcal{N}}\sigma^2_{\rm{n}}(e_iw_{\min}+c_i)(\alpha\sigma^4_it^{\rm{p}})^{-1}}$, and $E^*_{\rm{s}}(w_{\min})=2\sqrt{\sum_{i\in\mathcal{N}}\sigma^2_{\rm{n}}(e_iw_{\min}+c_i)t^{\rm{p}}(\alpha\sigma^4_i)^{-1}}+\sum_{i\in\mathcal{N}}(e_iw_{\min}+c_i)(\alpha N_{\rm{t}}\sigma^2_i)^{-1}$. For notation simplicity, we define here $A=\sum_{i\in\mathcal{N}}\sigma^2_{\rm{n}}e_it^{\rm{p}}(\alpha\sigma^4_{i})^{-1}$, $B=\sum_{i\in\mathcal{N}}\sigma^2_{\rm{n}}c_it^{\rm{p}}(\alpha\sigma^4_{i})^{-1}$, $C=\sum_{i\in\mathcal{N}}e_i(\alpha N_{\rm{t}}\sigma^2_i)^{-1}$, and
$D=\sum_{i\in\mathcal{N}}c_i(\alpha N_{\rm{t}}\sigma^2_i)^{-1}$. Then, $E^*_{\rm{s}}(w_{\min})=2\sqrt{Aw_{\min}+B}+Cw_{\min}+D$\,. The termination of Algorithm~\ref{Alg:PowerAllocation} gives us that the optimal $w_{\min}$ should satisfy that $E=E^*(w_{\min})=2\sqrt{Aw_{\min}+B}+Cw_{\min}+D$. This gives us the closed-form solution of $w_{\min}$ as
\begin{align*}
w_{\min}^*=\frac{(E-D)C+2A-\sqrt{[(E-D)C+2A]^2-C^2[(E-D)^2-4B]}}{C^2}\,.
\end{align*}
We can now derive the closed-form solution of $P^{\rm{p}}$ and $E_i^{\rm{t}}$. We observe that, in such an asymptotic case, the complexity of the approach is as low as $\mathcal{O}(N)$, which corresponds to the calculation of auxiliary variables $A$, $B$, $C$, and $D$.

\section{Numerical Results}\label{sec:numerical-results}
In this section, we numerically evaluate  the performance of Algorithm~\ref{Alg:PowerAllocation} to solve Problem~\eqref{Problem0} with non-linear and linear EH models. We use Matlab for performing numerical simulations. We first describe the set-ups of the simulations. Then, we test the convergence of the algorithm. Finally, we evaluate the average network sensing rates achieved by the algorithm with different network parameters.

\subsection{Simulation Set-ups}
The default set-ups of the simulations are given as follows. We deploy $N=20$ sensor nodes randomly in a disk region with radius $R$ meters. One BS is located at the centre of the region to transmit energy and to collect data. The BS has $N_{\rm{t}}=32$ antennas, and it has 3 Joules energy available in a time block of 1 second length. The frequency of the RF energy carrier is $915$~MHz. The path-loss depends on the distance between the BS and the node and is calculated according to the Friis equation. In the non-linear harvesting model, the RF-DC conversion rate function is $\eta(x)=P_{\max}(1-\exp{(-\eta_{\max} x/P_{\max}}))$, where $P_{\max}=20$~mW corresponds to the saturation of the RF-DC conversion power and $\eta_{\max}=0.3$ is the maximum RF-DC conversion rate. This non-linear model captures the saturation behaviour of the EH. For the linear model, we set the RF-DC conversion rate to be $\alpha=0.3$ by default. 

Each sensor node  transmits data at the standard $2.4$~GHz frequency. In each time block, the energy consumption to transmit a unit size data to the BS is $10^{-7}d^2$ Joules, where $d$ is the distance of the node to the BS. The static energy consumption of a node is $c_j=3\times 10^{-6}$ Joules. The time duration of channel estimation, $t^{\rm{p}}$, is 10\% of the time block. We use a simple least-square estimator to obtain CSI, with a noise level, $\sigma_{\rm{n}}^2$, at -90~dBm, and then apply maximum ratio transmit beamforming to send energy. The reason we use this PEB scheme is that we have a good closed-form approximation to the PEB gain $g(\cdot)$. However, it should be noticed that, the proposed Algorithm~\ref{Alg:PowerAllocation} can be applied to different PEB schemes, as long as the PEB gain meets the qualification condition of Assumption~\ref{assumpt:g}.

\subsection{Convergence Tests}

\begin{figure*}[t]
	\centering	
	\subfigure[][network sensing rate]{
		\includegraphics[width=0.43\textwidth]{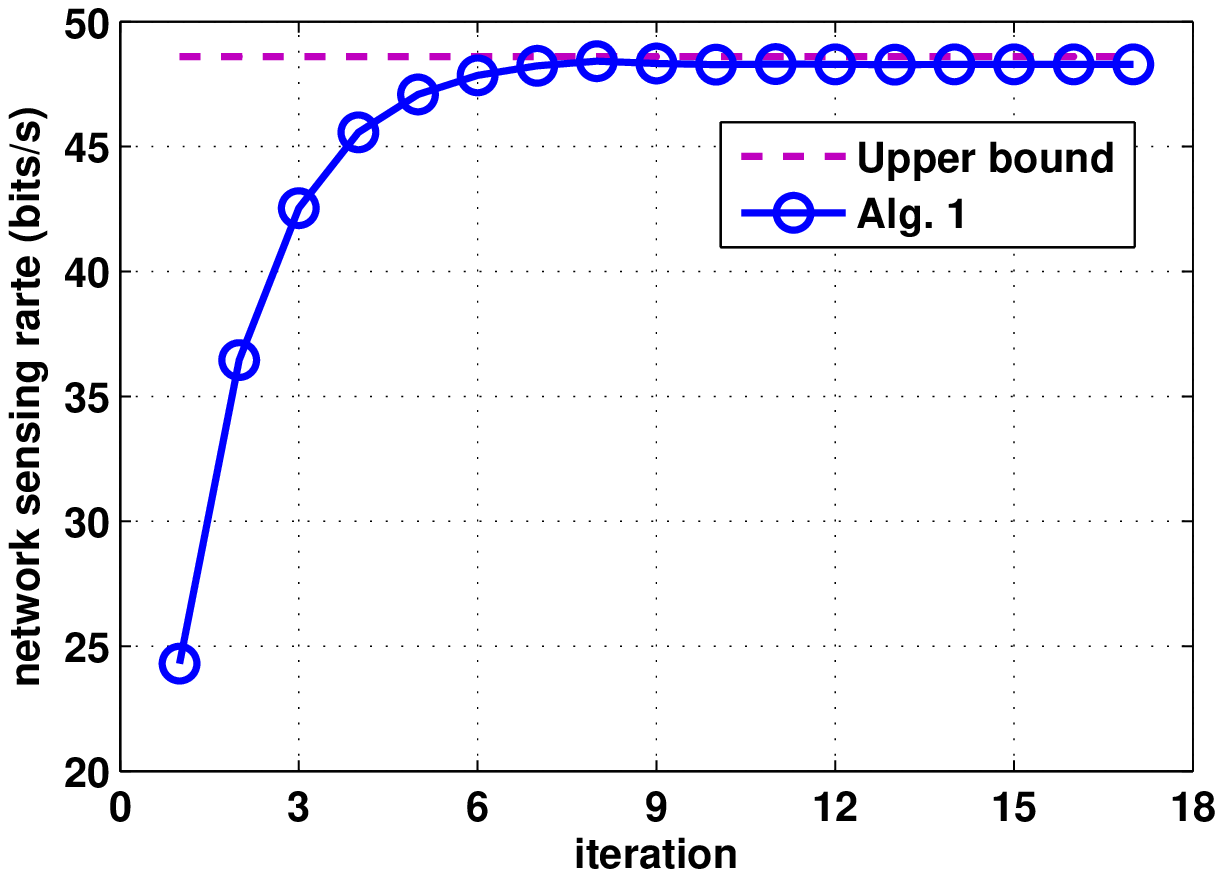}\label{Fig:Convergence_W_NonLinear}}
	\subfigure[][needed energy]{
		\includegraphics[width=0.43\textwidth]{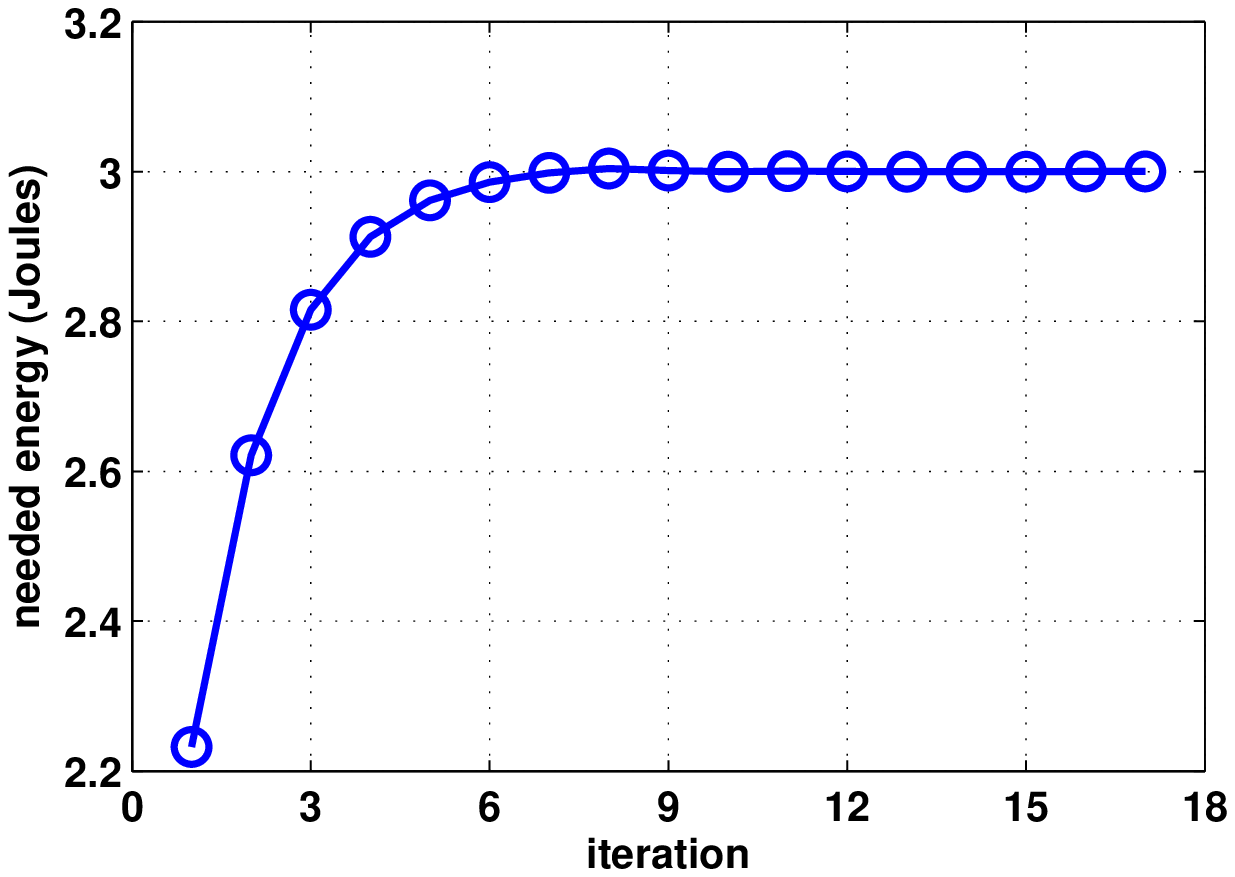}\label{Fig:Convergence_P_NonLinear}}
	\vspace{-0.2cm}
	\caption{Convergence of Algorithm~\ref{Alg:PowerAllocation}  (non-linear energy harvesting case).}\label{Fig:Convergence_NonLinear}
	\vspace{-0.8cm}
\end{figure*}

\subsubsection{Non-linear energy harvesting model}
First, we will show the convergence of Algorithm~\ref{Alg:PowerAllocation} for the non-linear EH model. The nodes are deployed in an annulus region with inner radius 25 meters and outer radius 50 meters. The termination parameter $\varepsilon$ in Line 6 of the algorithm  is set to be $0.001$ bit/s. Fig.~\ref{Fig:Convergence_NonLinear} shows $w_{\min}$ and the corresponding needed energy $E_{\textrm{s}}^*(w_{\min})$ achieved by Algorithm~\ref{Alg:PowerAllocation} in each iteration step. Recall that the optimal solution should satisfy that $E_{\textrm{s}}^*(w_{\min})=E$. We have that $E_{\textrm{s}}$ should converge to $E=3$ Joules. Initially, the sensing rate is $w=24.3$~bits/s, and the corresponding energy is $2.23$~Joules. Thus, in the second iteration, the threshold sensing rate increases to $36.46$~bits/s, and it keeps increasing until the 8th iteration, where the needed energy is slightly above $3$~Joules. Then, the threshold sensing rate starts decreasing. The algorithm terminates at the 17th iteration, where the resulting sensing rate is $48.29$~bits/s. The resulting energy is slightly less than $3$~Joules, which indicates the sensing rate is near optimal and feasible. The resulting sensing rate is close to the upper bound achieved according to Lemma~\ref{prop:wupperbound}, as shown by the purple dashed line. We can also see that the total number of iterations is not too large. Thus, the algorithm is efficient to achieve a near optimal solution.

\subsubsection{Linear energy harvesting model}

\begin{figure*}[t]
	\centering	
	\subfigure[][network sensing rate]{
		\includegraphics[width=0.43\textwidth]{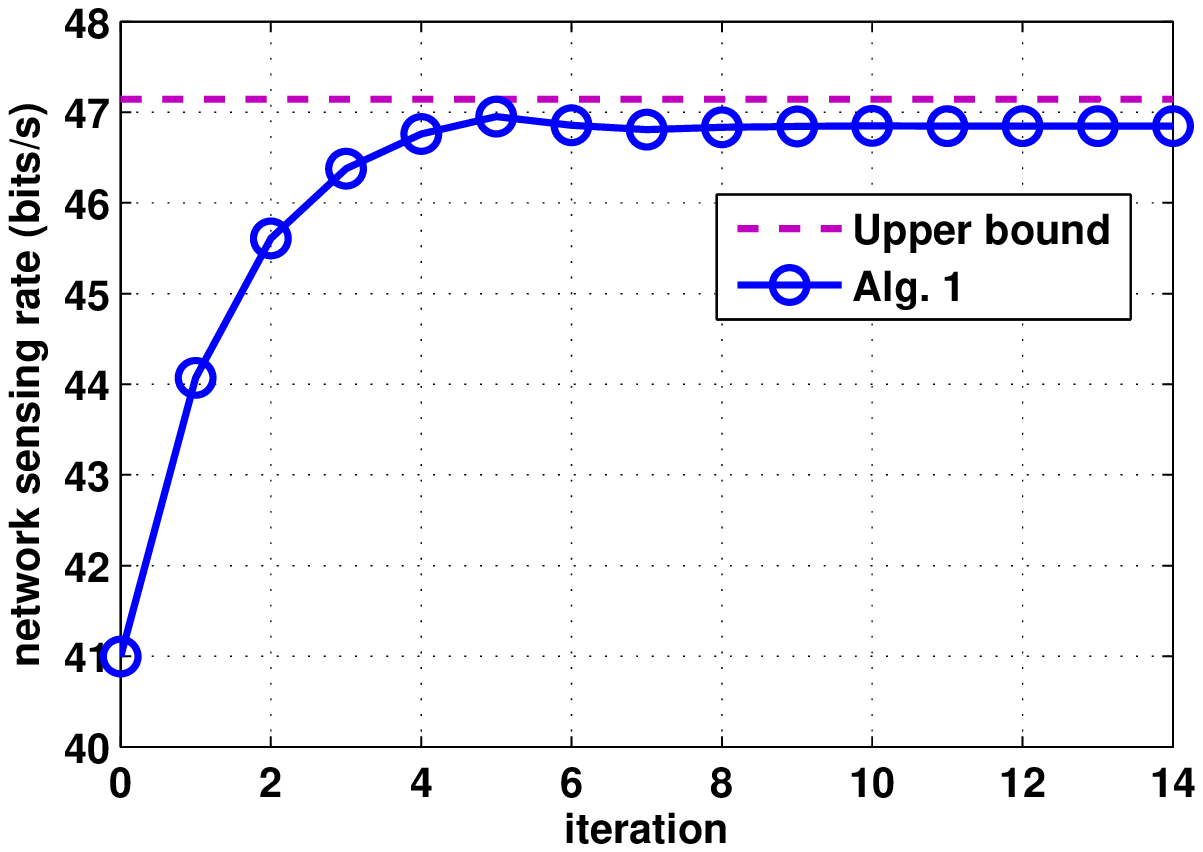}\label{Fig:Convergence_W}}
	\subfigure[][needed energy]{
		\includegraphics[width=0.43\textwidth]{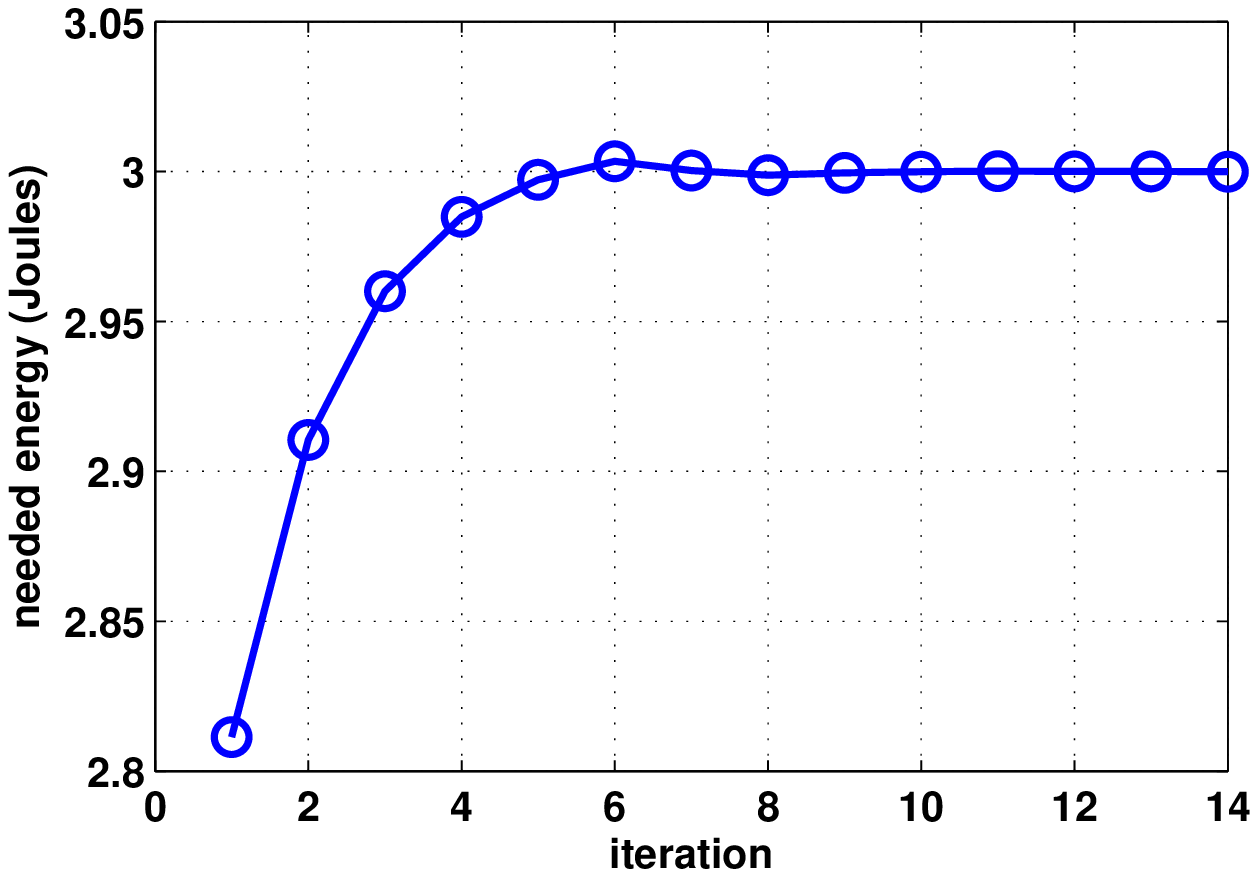}\label{Fig:Convergence_P}}
	\vspace{-0.2cm}
	\caption{Convergence of Algorithm~\ref{Alg:PowerAllocation} (linear energy harvesting case).}\label{Fig:Convergence}
	\vspace{-0.8cm}
\end{figure*}

Next, we show the convergence of Algorithm~\ref{Alg:PowerAllocation} in terms of network sensing rate for the problem with linear EH model. The convergence of the sensing rate and the required energy in a simulation case is shown in Fig.~\ref{Fig:Convergence}. The result is similar to the non-linear EH case in terms of convergence. More specifically,  the initial sensing rate is $w=41$~bits/s, and the corresponding energy is $2.811$~Joules. The sensing rate increases accordingly to $44.07$~bits/s until the 6th iteration, where the needed energy is slightly above $3$~Joules. Then, the threshold sensing rate starts to decrease. The algorithm terminates at the 15th iteration, where the resulting sensing rate is $46.85$~bits/s and it is close to the upper bound indicated by the purple dashed line. 

\subsection{Comparing Non-linear and Linear Models}
	\begin{figure}[t]
		\centering
		\includegraphics[width=0.5\textwidth]{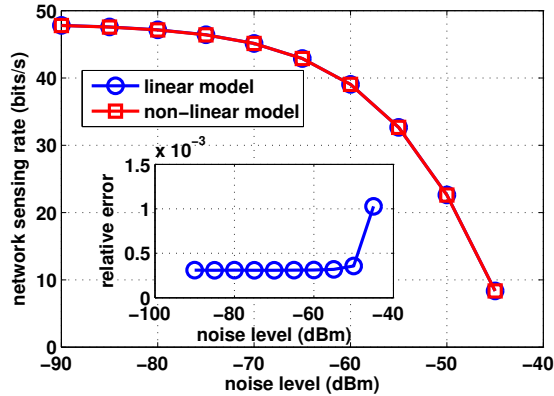}
			\vspace{-0.2cm}
		\caption{The network sensing rates achieved by Algorithm~\ref{Alg:PowerAllocation}, and the relative difference between the non-linear and linear model. }
		\label{Fig:Compare_Model}  
		\vspace{-0.8cm}
	\end{figure}

In this subsection, we will compare the results of Algorithm~\ref{Alg:PowerAllocation} for the problem with non-linear EH model and the one with linear EH model, to see whether the non-linear model we use here can be simplified to a linear model. We run simulation for 1000 times for different noise levels in channel estimation. In each simulation instance, we set the same network parameters, including the location of the nodes, the wireless channels, and the energy consumption model for the linear model and the non-linear model cases. The only difference is the EH model. We check the relative error of the results of using Algorithm~\ref{Alg:PowerAllocation} to solve the problem with the non-linear EH model and the ones with the linear model, i.e., $|w_{\min,\rm{NL}}-w_{\min,\rm{L}}|/w_{\min,\rm{NL}}$, where $w_{\min,\rm{NL}}$ and $w_{\min,\rm{L}}$ are the results for the non-linear and linear model respectively. The result is shown in Fig.~\ref{Fig:Compare_Model}. The blue line with circles and the red line with square marks represents the network sensing rate of the non-linear and the linear EH model, respectively, under different noise level in channel estimation. The two curves are close to each other, thus in the mini figure we show the relative error of the network sensing rate between the non-linear and linear model. We observe that, the relative error is less than $0.15\%$, which means that the solution for the linear case is close to the solution for the non-linear case. The main reason comes from the fact that, in our simulation, the optimal pilot power for the non-linear EH model cases lies in a region where the EH model has a good linearity, and far from saturation. If we use other non-linear EH models or different network set-ups, the difference may be large. However, investigating which EH models can be well approximated by a linear model is beyond the scope of this work. 

\subsection{Performance Tests}
To evaluate the performance of Algorithm~\ref{Alg:PowerAllocation}, we make simulations with different parameters, such as network size and noise level of channel estimation. For each combination of parameter, we simulate 1000 times with different deployments of the sensor nodes, and take the average.

The performance is compared to the upper bound achieved by Lemma~\ref{prop:wupperbound}, a sensing rate achieved by a random based power allocation, a rate achieved by a fixed power allocation~\cite{liu2014multi,yang2015throughput}, and a rate achieved by energy broadcasting, where there is no energy beamforming. The upper bound corresponds to the solution of Problem~\eqref{ProblemUpperbound_NonLinear}, which is the case where the BS has perfect CSI a priori. For the random based power allocation, the idea is that the BS first allocates the power for channel estimation $P^{\rm{p}}$ randomly, and then finds the solution of Problem~\eqref{Problem0} with $P^{\rm{p}}$ fixed. The fixed power allocation is similar to the random based power allocation, where the difference is that the BS always uses a fixed ratio of the total power (in the simulation we use 10\%) for channel estimation. Regarding the energy broadcasting case, the base station spends no power in channel estimation and just broadcasts energy with a fixed power of $3$~Watts. As default, we use purple dashed lines to represent the upper bound of the sensing rate, blue lines with circle marks for the sensing rate achieved by Algorithm~\ref{Alg:PowerAllocation}, red lines with square marks for the rate achieved by the fixed power allocation, green lines with crosses for the rate achieved by the random power allocation, and yellow lines with diamond marks for the rate achieved by energy broadcasting.

\subsubsection{Identical Expected Channel Estimation Gain} Since the linear model is a good approximation of the non-linear model, we consider the problem with linear model here for simplicity. Recall in Section~\ref{sec:LSEstimation}, we studied a special case where the BS uses LS Estimator for channel estimation. For the case where the expected channel estimation gains are identical, i.e., $\sigma_{h_i}^2=\sigma_{h_k}^2,\forall i\neq k$, we have a closed-form solution for Equation~\eqref{Eqn:OptPp_LS}, and it reduces the complexity of Algorithm~\ref{Alg:PowerAllocation} to $\mathcal{O}\left(N\log(E/(N\varepsilon))\right)$. Therefore, we first study the performance of Algorithm~\ref{Alg:PowerAllocation} in such an special case with different radius $R$, and different numbers of nodes $N$.

\begin{figure*}[t]
	\centering
	\subfigure[]{
		\includegraphics[width=0.43\textwidth]{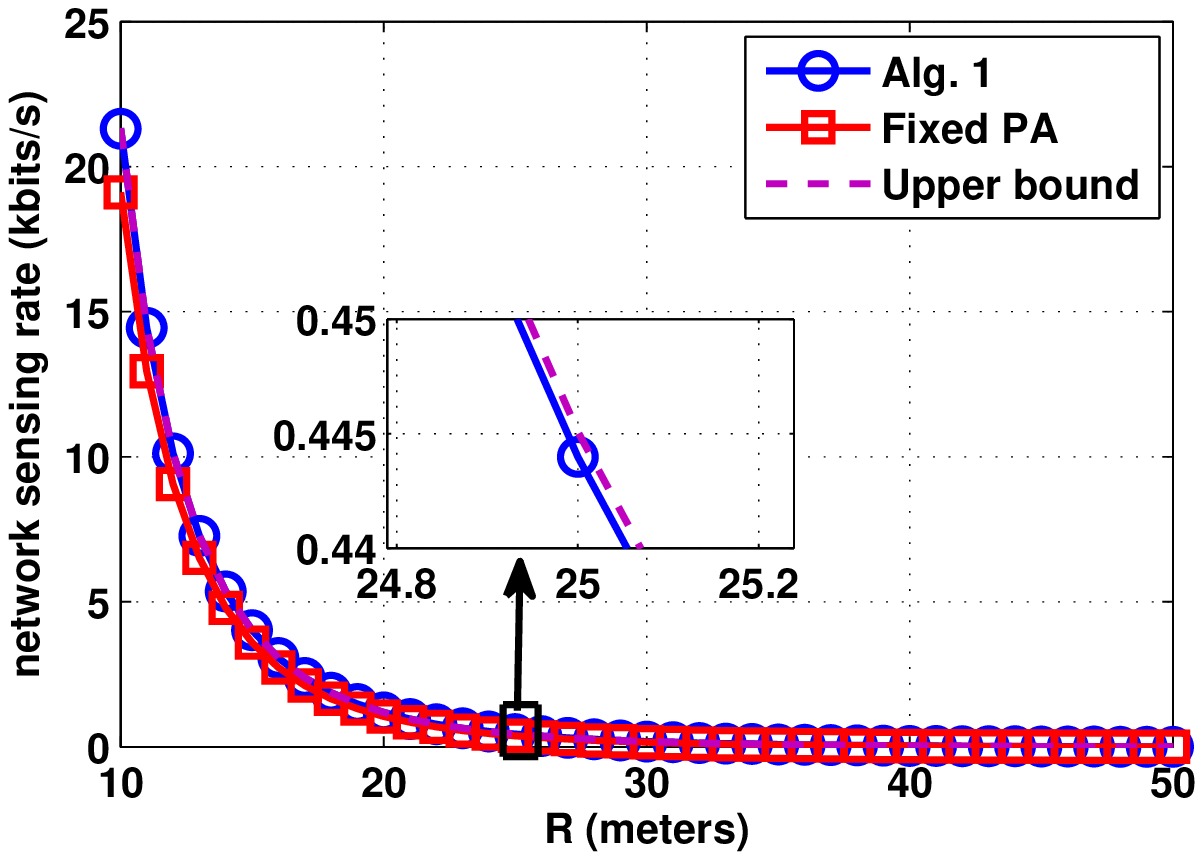}\label{Fig:R}}
	\subfigure[]{\includegraphics[width=0.43\textwidth]{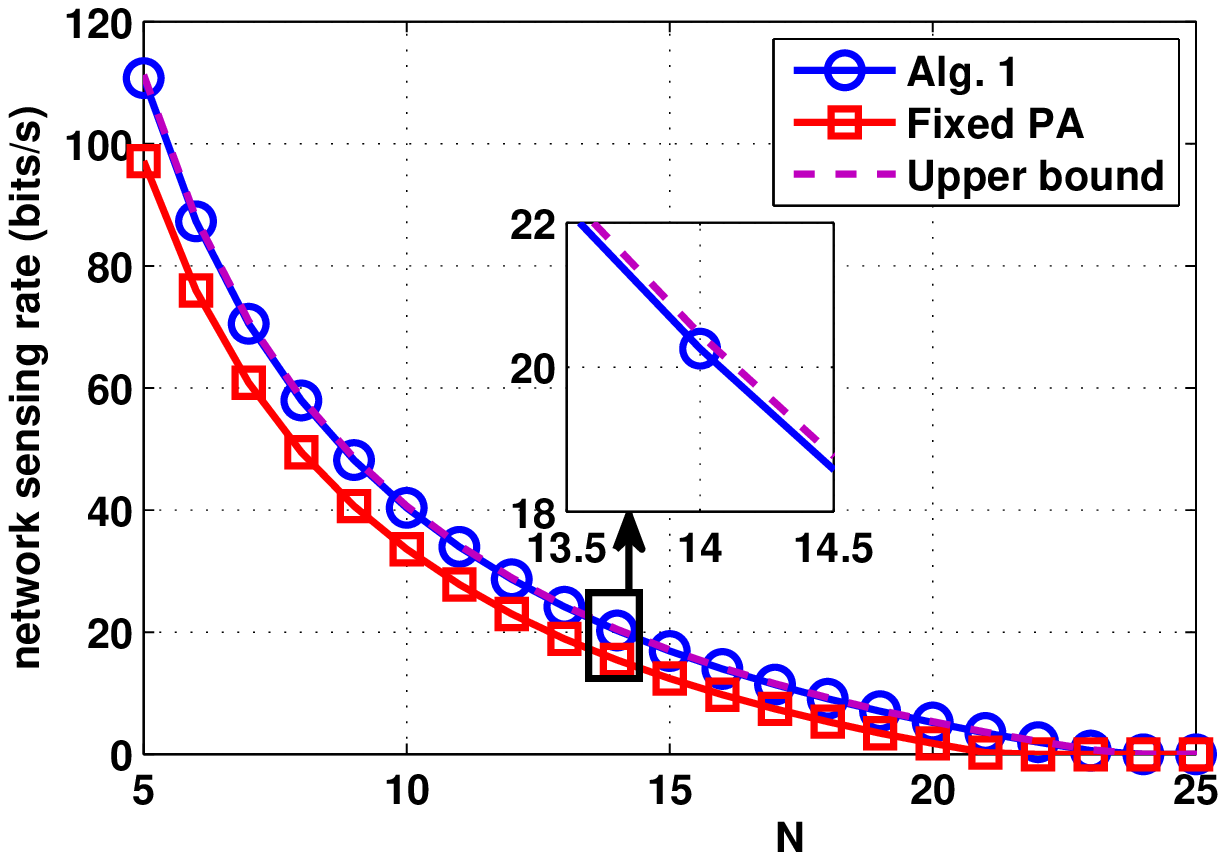}\label{Fig:N}}
	\vspace{-0.2cm}
	\caption{(a) Comparison of Algorithm~\ref{Alg:PowerAllocation} to other approaches with different radius $R$; (b) Comparison of Algorithm~\ref{Alg:PowerAllocation} to other approaches with different numbers of nodes $N$.}
	\vspace{-0.8cm}
\end{figure*}

For the case with different $R$, we vary it from 10 to 50. $N$ is fixed to be 20, and the distance of each node to the BS is $R$. The result is shown in Fig.~\ref{Fig:R}, where the X axis stands for the radius $R$, the Y axis stands for the network sensing rate, and the upper bound of the rate is achieved by Equation~\eqref{Eqn:wmin_Upperbound_LS}. In general, with larger distance from BS to the nodes, the energy received by each node drops significantly. Thus, the sensing rate decreases sharply.  We observe that, the rate achieved by energy broadcasting is 0 even when $R$ is 10 meters. This means that the energy received by the nodes is smaller than the constant energy consumption $c_i$. However, if we use energy beamforming, the nodes can harvest more energy then the constant energy consumption. When $R$ is 10~meters, the rate achieved by Algorithm~\ref{Alg:PowerAllocation} is approximately 21~kbits/s. Even when $R$ is 50 meters, the rate achieved by Algorithm~\ref{Alg:PowerAllocation} is 5~bits/s, which means that the power is still enough for the applications that require low sensing rate, such as agricultural monitoring, etc. Besides, the results show that, the rate achieved by Algorithm~\ref{Alg:PowerAllocation} is very close to the theoretical upper bound, as can be seen in the mini figure, and the relative difference is approximately $0.2\%$, which is small enough. Compared to the fixed power allocation, the rate achieved by~Algorithm~\ref{Alg:PowerAllocation} when $R=14$ is approximately $10\%$ higher, and it is approximately $30\%$ higher when $R=40$.

For the case with different $N$, we vary it from 5 to 25. The radius of the filed is $R=50$~meters. We compare the performance of Algorithm~\ref{Alg:PowerAllocation} to those achieved by the fixed power allocation, and the upper bound. The simulation results is shown in Fig.~\ref{Fig:N}. In general, the energy transmitted to each node, $E_i^{\rm{t}}$ is inversely proportional to $N$. Thus, when $N$ increases, each node receives less energy, and thus the sensing rate reduces. The rate achieves by Algorithm~\ref{Alg:PowerAllocation} is close to the theoretical upper bound, and it is approximately $10\%$ larger than the one achieved by the fix power allocation. When $N$ is larger than 24, the harvested energy of the nodes is smaller than $c_i$, and the sensing rate becomes 0. We can conclude that, the proposed Algorithm~\ref{Alg:PowerAllocation} achieves a near optimal solution, and the upper bound is tight when the noise level for channel estimation is low. Also, with the power allocation achieved from Algorithm~\ref{Alg:PowerAllocation}, the WPSN with even one BS that provides energy can monitor a region of size of several hundred meters square with sensing rate at approximately several bits per second.

\subsubsection{General Cases}
Now we will consider the general cases where the channel estimation gain is not necessarily identical for all the nodes. The nodes are randomly deployed in a disk region with $R=50$~meters. 

\begin{figure*}[t]
		\centering
		\subfigure[]{
			\includegraphics[width=0.43\textwidth]{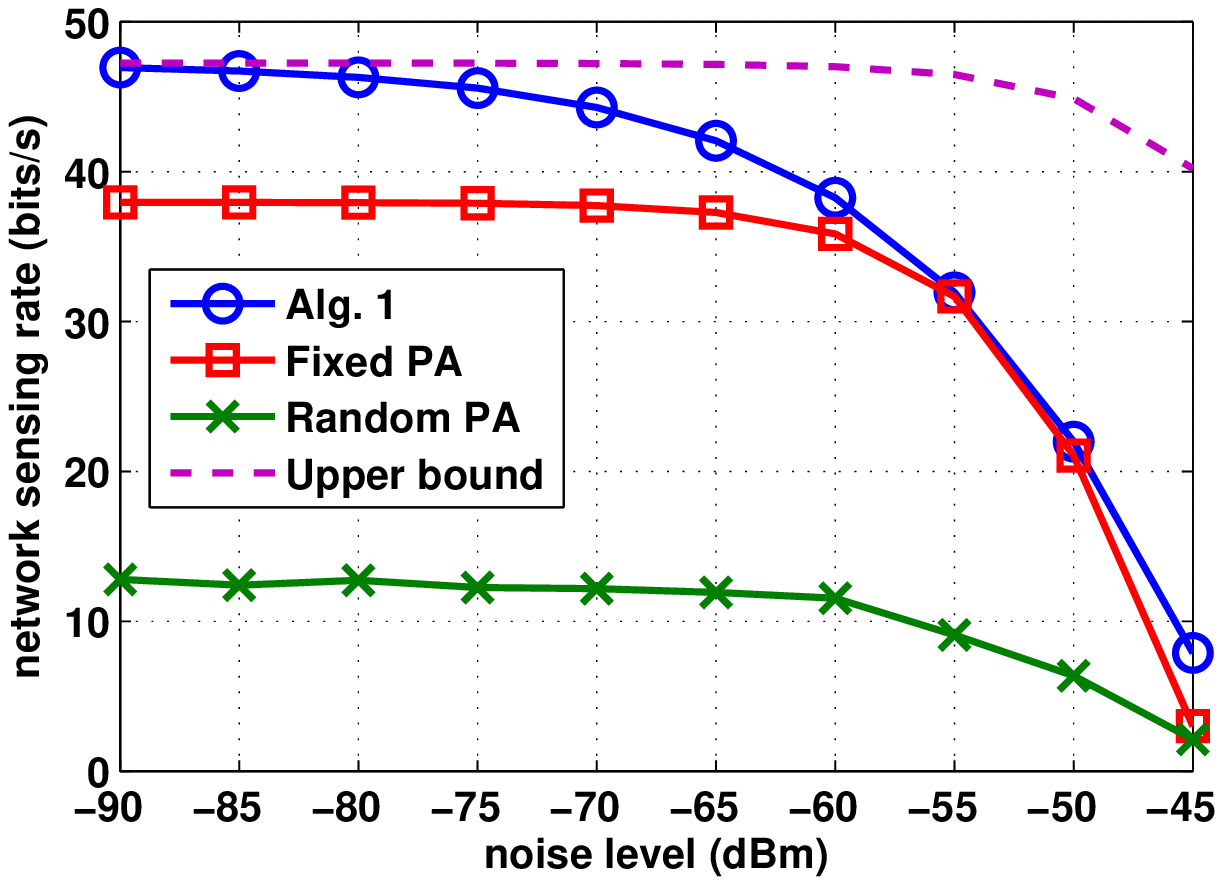}\label{Fig:Noise}}
			\subfigure[]{\includegraphics[width=0.43\textwidth]{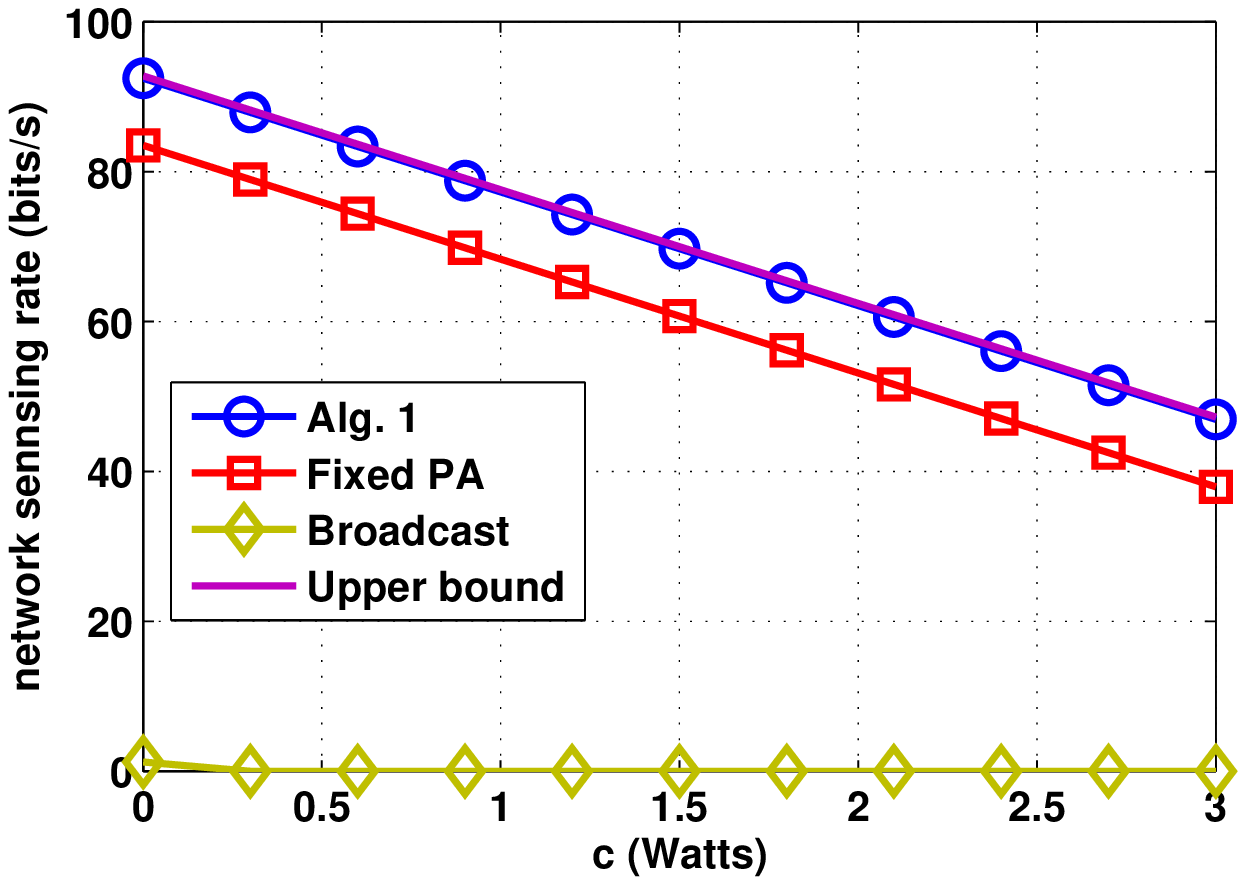}\label{Fig:c}}
		\vspace{-0.2cm}
		\caption{(a) Comparison of Algorithm~\ref{Alg:PowerAllocation} to other approaches with different noise levels; (b) Comparison of Algorithm~\ref{Alg:PowerAllocation} to other approaches with different static power consumptions $c$.}
		\vspace{-0.8cm}
	\end{figure*}

To begin with, we test Algorithm~\ref{Alg:PowerAllocation} with different noise levels. For the same node deployment and channel state, we vary the noise level from $-90$~dBm to $-45$~dBm, and compare the network sensing rate achieved by different algorithms. With lower noise level, the channel estimation is more power efficient. The simulation result is shown in Fig.~\ref{Fig:Noise}. In general, with higher noise level, the channel estimation error becomes larger. Thus, the BS tends to spend more energy in channel estimation, resulting in less energy that is transmitted to the nodes. Therefore, the sensing rate that is achieved by Algorithm~\ref{Alg:PowerAllocation} decreases from $47$~bits/s to $7$~bits/s when the noise level increases from $-90$~dBm to $-45$~dBm.  By comparing the performance to the upper bound, we can see that the gap between the sensing rate achieved by Algorithm~\ref{Alg:PowerAllocation} and the upper bound becomes larger when the noise level is high. Therefore, the upper bound is loose at high noise level. For the performance of fixed power allocation, it can be observed that, the sensing rate is close to the one achieved by Algorithm~\ref{Alg:PowerAllocation} when the noise level is around $-55$ to $-50$~dBm. It means that, in these cases the pilot power of the fixed power scheme, $0.3$~Watts, happens to be close to the optimal one. However, for other noise level, it is much worse than Algorithm~\ref{Alg:PowerAllocation}. The rate drops to approximately $3$~bits/s when the noise level is at $-45$~dBm. Besides, the sensing rate achieved by random power allocation is even worse than the fixed power allocation. It is approximately only one fourth to the rate achieved by Algorithm~\ref{Alg:PowerAllocation}.
	
Next, we also compare the sensing rate achieved by different algorithms with different static power consumption $c$, as shown in Fig.~\ref{Fig:c}. The noise level is $-90$~dBm. It can be seen that, the network sensing rate decreases linearly with $c$, until when it reaches 0. When $c=0$, the rate achieved by energy broadcasting is approximately 1.2 bits/s. However, when $c$ is slightly larger than 0, the harvested energy from energy broadcasting can not support the static power consumption, and thus the nodes that are far away from the BS have no residual power to transmit the data. With energy beamforming, the BS can better allocate the power to different nodes. Therefore, it may transmit more energy to the nodes that are further away than the nearby nodes, and the network sensing rate is much larger than the case by energy broadcasting. We observe that, under different $c$, the network sensing rate is always close to the upper bound, and it is basically linearly decreasing with $c$.

To summarize, the simulation results show the convergence of Algorithm~\ref{Alg:PowerAllocation}. Its performance is close to the upper bound if we have large power efficiency in channel estimation. Also, Algorithm~\ref{Alg:PowerAllocation} outperforms other power allocations scheme in terms of the sensing rate.

\section{Conclusions and Future Works}\label{sec:conclusions}
%

We studied a wirelessly-powered sensor network where a BS with multiple antennas needs to optimally allocate its power in pilot transmission, and in energy beamforming to a set of sensor nodes. We considered a non-linear EH of the nodes, and represented the model by a generic class of non-linear functions. We showed that this optimal power allocation is a non-convex optimization problem. We proposed an algorithm to achieve a near optimal solution of the problem. We studied the convergence properties of the proposed algorithm and showed in the simulation that the solutions achieved by the proposed algorithm are close to the theoretical upper-bound. Moreover, the solutions improve the network sensing rate by around 10\% compared to the existing fixed power allocation approach. 

In the future, we will study the case where the channel states are time-correlated, and thus the base station can update the power allocation adaptively. Another interesting topic is to jointly consider the data routing decision in the problem. We are also interested in the cases with multiple BSs, where the BSs may perform cooperative channel estimation and beamforming.
\appendices
\section{Proofs}
\label{appendix:proofs}
\subsection{Proposition~\ref{prop:convex}}
Recall that function $g_i$ is monotone increasing and positive, and its first derivative is monotone decreasing. Recall that
$f_i(P^{\rm{p}};w_{\min})\triangleq\eta^{-1}(e_iw_{\min}+c_i)/(g_i(P^{\rm{p}}))$.  Thus, we have that
$f_i>0$, 
\begin{align*}
\frac{d f_i(P^{\rm{p}};w_{\min})}{d P^{\rm{p}}}=-\eta^{-1}(e_iw_{\min}+c_i)\frac{g_i'(P^{\rm{p}})}{g_i^2(P^{\rm{p}})}\leq 0\,,
\end{align*}
and
\begin{align*}
\frac{d^2 f_i(P^{\rm{p}};w_{\min})}{d (P^{\rm{p}})^2}=\eta^{-1}(e_iw_{\min}+c_i)\frac{2\left(g_i'(P^{\rm{p}})\right)^2-g_i''(P^{\rm{p}})g_i^2(P^{\rm{p}})}{g_i^4(P^{\rm{p}})}\geq 0\,,
\end{align*}
i.e., $f_i$ is monotone decreasing, positive, and its first derivative is monotone increasing w.r.t. $P^{\rm{p}}$. Thus, $E_s(P^{\rm{p}}|w_{\min})$ is convex, and the problem is a convex optimization problem. Thus, we have that
\begin{align*}
\frac{d E_{\rm{s}}(P^{\rm{p}}|w_{\min})}{d P^{\rm{p}}}=t^{\rm{p}}+\sum_{i\in\mathcal{N}}f_i'(P^{\rm{p}}|w_{\min})\,.
\end{align*}
Then, we have that the first derivative of $E_{\rm{s}}(P^{\rm{p}}|w_{\min})$ satisfies one of the following cases: 1) it is always positive; 2) it has a root; 3) it is always negative.

The first case gives us that the optimal solution is $P^{\rm{p}}=0$. The second case gives us that the optimal solution $P^{\rm{p}}$ satisfies
\begin{align}
\label{Eqn:OptPp}
\sum_{i\in\mathcal{N}}f_i'(P^{\rm{p}}|w_{\min})=-t^{\rm{p}}\,.
\end{align}
The third case means the optimal $P^{\rm{p}}\rightarrow +\infty$, which means that the given $w_{\min}$ is not achievable and thus we discard this case. To conclude, given $w_{\min}$, the optimal training power $P^{\rm{p}}$ such that $E$ is minimized, is either $P^{\rm{p}}=0$, or it satisfies Equation~\eqref{Eqn:OptPp}. This completes the proof.

\subsection{Lemma~\ref{prop:wupperbound}}
We start the proof with comparing Problem~\eqref{Problem0} and Problem~\eqref{ProblemUpperbound_NonLinear}. Recall that $\eta_{\max}\triangleq\max_{x\geq 0}\eta(x)/x$. We have that the right hand side of Constraint~\eqref{ProblemUpperbound_NonLinear:C1}, $\eta_{\max} E_i^{\rm{t}} g_{i}\left(E/t^{\rm{p}}\right)$ is larger than or equal to $\eta(E_i^{\rm{t}} g_{i}\left(E/t^{\rm{p}}\right))$. Together with Constraint~\eqref{Problem0:Psum}, we have that $\eta_{\max} E_i^{\rm{t}} g_{i}\left(E/t^{\rm{p}}\right)$ is larger than or equal to $\eta(E_i^{\rm{t}} g_{i}\left(P^{\rm{p}}\right))$, which is the right hand side of Constraint~\eqref{Problem0:energy}. Therefore, Constraint~\eqref{ProblemUpperbound_NonLinear:C1} is a relaxation of Constraint~\eqref{Problem0:energy}. Additionally, we have that Constraint~\eqref{ProblemUpperbound_NonLinear:C2} is a relaxation of Constraint~\eqref{Problem0:Psum}. Therefore, we have that Problem~\eqref{ProblemUpperbound_NonLinear} is a relaxation of Problem~\eqref{Problem0}. Since the problems are maximization problems, we have that the optimum of Problem~\eqref{Problem0} is upper bounded by the optimum of Problem~\eqref{ProblemUpperbound_NonLinear}.

Next, we will find an upper bound of the optimum of Problem~\eqref{ProblemUpperbound_NonLinear}.
Briefly speaking, Constraint~\eqref{ProblemUpperbound_NonLinear:C1} gives us that the optimal $w_{\min}$ should satisfy $(e_iw_{\min}+c_i)/(\eta_{\max}g_i(E/t^{\rm{p}})), \forall i\in\mathcal{N}$. 
 Together with Constraint~\eqref{ProblemUpperbound_NonLinear:C2}, we have that 
\begin{align*}
\sum_{i\in\mathcal{N}}\frac{e_i}{g_i(\frac{E}{t^{\rm{p}}})}w_{\min}+\sum_{i\in\mathcal{N}}\frac{c_i}{g_i(\frac{E}{t^{\rm{p}}})}=\eta_{\max} E\,,
\end{align*}
whose solution gives us the right hand side of Equation~\eqref{Eqn:wupperbound}. Thus, we have that such an upper bound of the optimum of Problem~\eqref{ProblemUpperbound_NonLinear} is also an upper bound of that of Problem~\eqref{Problem0}, which completes the proof.

\subsection{Proposition~\ref{thm:Th1}}
Denote $w_{\min}^{\rm{o}}, \vec{E}^{\rm{t,o}}, P^{\rm{p,o}}$ the optimal solution of Problem~\eqref{Problem0}. In the initialization of Algorithm~\ref{Alg:PowerAllocation}, if Problem~\eqref{Problem0} is feasible, then we have that $w_{\min}^{\rm{l}}=0\leq w_{\min}^{\rm{o}}<w_{\min}^{\rm{u}}$. We will prove that, after each iteration of Algorithm~\ref{Alg:PowerAllocation}, $w_{\min}^{\rm{o}}$ always lies within the range $[w_{\min}^{\rm{l}}, w_{\min}^{\rm{u}})$. This is equivalent to prove that, for a $w_{\min}$, if $E_{\rm{s}}^*(w_{\min})-E>0$, then $w_{\min}>w_{\min}^{\rm{o}}$; otherwise, $w_{\min}\leq w_{\min}^{\rm{o}}$.

We prove that if $E_{\rm{s}}^*(w_{\min})-E>0$, then $w_{\min}>w_{\min}^{\rm{o}}$, by contradiction. Assume that in this case $w_{\min}\leq w_{\min}^{\rm{o}}$. Since $w_{\min}^{\rm{o}}$ is feasible. We have that $E\geq E_{\rm{s}}^*(w_{\min}^{\rm{o}})=t^{\rm{p}}P^{\rm{p}}(w_{\min}^{\rm{o}})+\sum_if_i(P^{\rm{p}}(w_{\min}^{\rm{o}});w_{\min}^{\rm{o}})$, where $P^{\rm{p}}(w_{\min}^{\rm{o}})$ is the optimal $P^{\rm{p}}$ for Problem~\eqref{ProblemSub_NonLinear} when problem input $w_{\min}$ is $w_{\min}^{\rm{o}}$. Recall that function $f_i(P^{\rm{p}};w)$ is monotone increasing with $w$. Thus, $f_i(P^{\rm{p}}(w_{\min}^{\rm{o}});w_{\min}^{\rm{o}})\geq f_i(P^{\rm{p}}(w_{\min}^{\rm{o}});w_{\min})$. Therefore, $E_{\rm{s}}^*(w_{\min}^{\rm{o}})=t^{\rm{p}}P^{\rm{p}}(w_{\min}^{\rm{o}})+\sum_if_i(P^{\rm{p}}(w_{\min}^{\rm{o}});w_{\min}^{\rm{o}}) \geq t^{\rm{p}}P^{\rm{p}}(w_{\min}^{\rm{o}})+\sum_if_i(P^{\rm{p}}(w_{\min}^{\rm{o}});w_{\min})\geq P^{\rm{p}}(w_{\min})+\sum_if_i(P^{\rm{p}}(w_{\min});w_{\min})=E_{\rm{s}}^*(w_{\min})>E$, which contradicts to that $E^*(w_{\min}^{\rm{o}})\leq E$.

Similarly, we can prove that if $E_{\rm{s}}^*(w_{\min})-E\leq 0$, then $w_{\min}\leq w_{\min}^{\rm{o}}$. Thus, we can conclude that after each iteration of Algorithm~\ref{Alg:PowerAllocation}, $w_{\min}^{\rm{o}}$ always lies within the range $[w_{\min}^{\rm{l}}, w_{\min}^{\rm{u}})$. Furthermore, since $w_{\min}^{\rm{l}}$ is always feasible in each iteration, and we set the $w_{\min}$ to be $w_{\min}^{\rm{l}}$, we have that the output of the algorithm is feasible. Moreover, since the algorithm terminates when $w_{\min}^{\rm{u}}-w_{\min}^{\rm{l}}\leq \varepsilon$, we have that $\varepsilon\geq w_{\min}^{\rm{o}}-w_{\min}^{\rm{l}}= w_{\min}^{\rm{o}}-w_{\min}$, where the equality comes from Line 19 of the algorithm. This  completes the proof.

\subsection{Proposition~\ref{prop:time-complexity}}
Recall that the algorithm is based on bisection search. Therefore, the number of iterations depends on the initial range, which is $[0,w_{\min}^{\rm{u}}]$, and the termination condition $\varepsilon$. Recall that the initial $w_{\min}^{\rm{u}}$ is the optimal solution of Problem~\eqref{ProblemUpperbound_NonLinear}. We can achieve an upper bound of $w_{\min}^{\rm{u}}$ as $E\eta_{\max}/\left(N\min\{e_i/g_i\}\right)$, which is $\mathcal{O}(E/N)$. Therefore, the number of iteration is $\mathcal{O}(\log(E/(N\varepsilon)))$. In each iteration, the algorithm achieves $E_{\rm{s}}^*(w_{\min})$ according to Proposition~\ref{prop:convex}, whose complexity is $\mathcal{O}(N\log(E))$. Thus, the total complexity is $\mathcal{O}\left(N\log(E)\log(E/(N\varepsilon))\right)$.
\subsection{Remark~\ref{rem:asymptotic_g}}
We denote $\alpha=P^{\rm{p}}/N_{\rm{t}}$. Then, we have that when $N_{\rm{t}}$ is fixed, the monotonicity and concavity of a function of $P^{\rm{p}}$ is the same to that of $\alpha$. Here, we will analyze how $g_i(P^{\rm{p}})$ behaves when $N_{\rm{t}}$ is sufficiently large. From Equation~\eqref{Eqn:g_j} we have that
\begin{subequations}
\begin{align}
\label{Eqn:g_j:Asympt_1}
\lim_{N_{\rm{t}}\rightarrow +\infty} g_i({P^{\rm{p}}})&=\lim_{N_{\rm{t}}\rightarrow +\infty} g_i({\alpha N_{\rm{t}}})\\
&=\lim_{N_{\rm{t}}\rightarrow +\infty}\mathbb{E}\left[\frac{\alpha N_{\rm{t}}\vec{h}_i^H\vec{h}_i\vec{h}_i^H\vec{h}_i+2\sqrt{\alpha}N_{\rm{t}}\vec{n}_i^H\vec{h}_i\vec{h}_i^H\vec{h}_i+N_{\rm{t}}\vec{n}_i^H\vec{h}_i\vec{h}_i^H\vec{n}_i}{\alpha N_{\rm{t}}\vec{h}_i^H\vec{h}_i+2\sqrt{\alpha}N_{\rm{t}}\vec{h}_i^H\vec{n}_i+N_{\rm{t}}\vec{n}_i^H\vec{n}_i}\right]\\
&=\mathbb{E}\left[\lim_{N_{\rm{t}}\rightarrow +\infty}\frac{\alpha\frac{\vec{h}_i^H\vec{h}_i\vec{h}_i^H\vec{h}_i}{N^2_{\rm{t}}}+2\sqrt{\alpha}\frac{\vec{n}_i^H\vec{h}_i\vec{h}_i^H\vec{h}_i}{N^2_{\rm{t}}}+\frac{\vec{n}_i^H\vec{h}_i\vec{h}_i^H\vec{n}_i}{N^2_{\rm{t}}}}{\frac{\alpha \vec{h}_i^H\vec{h}_i}{N_{\rm{t}}}+2\sqrt{\alpha}\frac{\vec{h}_i^H\vec{n}_i}{N_{\rm{t}}}+\frac{\vec{n}_i^H\vec{n}_i}{N_{\rm{t}}}}N_{\rm{t}}\right]\,.
\end{align}
\end{subequations}
Recall that $\vec{h}_i=[h_{i1},\ldots,h_{iN_{\rm{t}}}]^T$, and the element of $\vec{h}_{i}$ are i.i.d.. Thus, according to the random matrix theory~\cite{tulino2004random}, we have that $\lim_{N_{\rm{t}}\rightarrow +\infty}\vec{h}_i^H\vec{h}_i/N_{\rm{t}}\xrightarrow{\text{a.s.}} \sigma_{i}^2$. Similarly, we have that $ \lim_{N_{\rm{t}}\rightarrow +\infty}\vec{n}_i^H\vec{n}_i/N_{\rm{t}}\xrightarrow{\text{a.s.}} \sigma_{\rm{n}}^2$. Since $\vec{h}_i$ and $\vec{n}_i$ are mutually independent, we have that $\lim_{N_{\rm{t}}\rightarrow +\infty}\vec{h}_i^H\vec{n}_i/N_{\rm{t}}\xrightarrow{\text{a.s.}} 0$. Thus, we can continue to simplify Equation~\eqref{Eqn:g_j:Asympt_1} as follows:
\begin{align*}
\lim_{N_{\rm{t}}\rightarrow +\infty} g_i({P^{\rm{p}}})&\xrightarrow{\text{a.s.}}\mathbb{E}\left[\lim_{N_{\rm{t}}\rightarrow +\infty}N_{\rm{t}}\frac{\alpha\sigma^4_{i}}{\alpha\sigma^2_{i}+\sigma^2_{\rm{n}}}\right]=\frac{\alpha \sigma^4_{i}}{\alpha \sigma^2_{i}+\sigma^2_{\rm{n}}}\lim_{N_{\rm{t}}\rightarrow +\infty}N_{\rm{t}}\,.
\end{align*}
By replacing $\alpha $ by $P^{\rm{p}}/N_{\rm{t}}$, we have that $g_i(P^{\rm{p}})$ converges almost surely to $N_{\rm{t}}\sigma_i^4P^{\rm{p}}(\sigma^2_iP^{\rm{p}}+N_{\rm{t}}\sigma^2_{\rm{n}})^{-1}$. One can easily check that the function is monotone increasing and concave w.r.t. $P^{\rm{p}}$  by checking its derivatives. This completes the proof.

\section{Discussion on the feasibility of Assumption~\ref{assumption:eta} for adaptive waveform design}
\label{appendix:adapative-waveform}
When the BS can transmit energy with different sinewaves, it could allocate power to different sinewaves according to the CSI. This scheme is called adaptive waveform deign, and is shown to be better than non-adaptive ones in terms of harvested energy in~\cite{clerckx2016waveform}, especially under the non-linear RF-DC model. Here, we will show that, using some adaptive waveform design schemes, the energy transmission using multisine carriers and the non-linear RF-DC model will satisfy our Assumption~\ref{assumption:eta}.

To illustrate the idea, we reuse the toy-example in~\cite[Section IV.B]{clerckx2016waveform}: There are two sinewaves, with $s_0>0$, $s_1>0$ be the amplitude of the sinewaves, and $A_0>0$, $A_1>0$ the channels amplitude of the two sinewaves. Reference~\cite{clerckx2016waveform} shows that, to maximizing the output DC is (approximately) equivalent to maximizing
\begin{align*}
z_{DC}(s_0,s_1)=\tilde{k}_2(s_0^2A_0^2+s_1^2A_1^2)+\tilde{k}_4[(s_0^2A_0^2+s_1^2A_1^2)^2+2s_0^2s_1^2A_0^2A_1^2]\,,
\end{align*}
where $\tilde{k}_2$ and $\tilde{k}_4$ are two positive coefficients that are related to the RF-DC circuit.

Our Assumption 2 requires that, if the received RF power at the antenna, $P_{ave,rf}$, increases to $P_{ave,rf}'> P_{ave,rf}$, the harvested power increases, and so as the corresponding $z_{DC}(s_0, s_1)$, where $P_{ave,rf}=0.5(s_0^2A_0^2+S_1^2A_1^2)$. This may not necessarily be true in general, though there exist some waveform design schemes for which this assumption holds: 
\begin{enumerate}
	\item Allocates power to the strongest frequency component with adaptive single sinewave (ASS) strategy in~\cite{clerckx2016waveform}: Without loss of generality, assume that $A_0>A_1$, i.e., frequency 0 has stronger amplitude than frequency 1. Then, $P_{ave,rf}'> P_{ave,rf}$ indicates that $s_0'>s_0$ and $s_1=s_1'=0$. This yields $z_{DC}(s_0,s_1)<z_{DC}(s'_0,s'_1)$, i.e., the harvested power increases.
	\item Allocating same power ratio to two frequency components, i.e., $s'_0/s'_1=s_0/s_1$: If the ratio depends on the channel strength, then it is the adaptive matched filter in~\cite{clerckx2016waveform}. Otherwise, it is a non-adaptive scheme (e.g. it is the uniform power scheme in [38] if the ratio is 1). In this case, $P'_{ave,rf}> P_{ave,rf}$ indicates that $s'_0>s_0$ and $s_1>s'_1$ (since $s'_0/s'_1=s_0/s_1$). This again yields $z_{DC}(s_0,s_1)<z_{DC}(s'_0,s'_1)$.
	\item Optimal scheme that maximizes $z_{DC}$: In this case, $(s_0,s_1)=\arg\max_{s_0,s_1} \left\{z_{DC}(s_0,s_1)| s_0^2A_0^2\right.$ $\left.+S_1^2A_1^2= 2P_{ave,rf} \right\}$, and $(s'_0,s'_1)=\arg\max_{s_0,s_1}\left\{z_{DC}(s_0,s_1)| s_0^2A_0^2+S_1^2A_1^2= 2P'_{ave,rf} \right\}$. Since $P'_{ave,rf}> P_{ave,rf}$, we can construct a feasible point $(s''_0,s_1)$ by setting 
	\begin{align*}
	s''_0= \sqrt{\frac{2(P'_{ave,rf}-P_{ave,rf})}{A_0^2}+s_0^2}\,,
	\end{align*}
	such that $(s''_0)^2A_0^2+s_1^2A_1^2= 2P'_{ave,rf}$ and $s''_0>s_0$. We have that $z_{DC}(s''_0,s_1)>z_{DC}(s_0,s_1)$. Meanwhile, since $(s''_0,s_1)$ is a feasible point of $\arg\max_{s_0,s_1}\left\{z_{DC}(s_0,s_1)| s_0^2A_0^2+S_1^2A_1^2= 2P'_{ave,rf} \right\}$, we have that $z_{DC}(s''_0,s_1)\leq z_{DC}(s'_0,s'_1)$. Thus again, $z_{DC}(s_0,s_1)< z_{DC}(s'_0,s'_1)$.
\end{enumerate}
These examples show that, the non-linear RF-DC model with some adaptive but suboptimal waveform design schemes could still satisfy Assumption 2. It also works for some non-adaptive schemes. However, there are some other waveform design schemes that may fail, such as randomly assigning the power to different frequencies (which is non-adaptive and suboptimal), and some adaptive but suboptimal schemes (for which we could not prove that $z_{DC}(s_0,s_1)< z_{DC}(s'_0,s'_1)$ when $P_{ave,rf}< P'_{ave,rf}$).

\section{Concavity of PEB gain for LS cases}
\label{appendix:concavity}
We study the concavity of the PEB gain $g_i(P^{\rm{p}})$ in~\eqref{Eqn:g_j} and achieve the following result:
\begin{prop}
$g_i(P^{\rm{p}})$ in Equation~\eqref{Eqn:g_j} is concave almost for sure in the region $P^{\rm{p}}\geq (2\sqrt{3}-1)Q^{-1}(N_{\rm{t}},0,0.99)(N_{\rm{t}}\sigma^2_{\rm{n}}/\vec{h}_i^H\vec{h}_i)$, where $2Q^{-1}(N_{\rm{t}},0,0.99)$  is the inverse of the generalized regularized incomplete gamma function that represents the $99$-th percentile from the lower cumulative distribution of the Chi-squared distribution with degree of freedom $2N_{\rm{t}}$ (corresponding to the real parts and the imaginary parts of the noises), i.e., the probability that $\|\vec{n}\|^2\leq Q^{-1}(N_{\rm{t}},0,0.99)\sigma_{\rm{n}}^2$ is $99$\%.
\end{prop}
\begin{IEEEproof}
Recall that $g_i(P^{\rm{p}})=\mathbb{E}\left[\|\sqrt{P^{\rm{p}}}\vec{h}_i^H\vec{h}_i+\sqrt{N_{\rm{t}}}\vec{h}_i^H\vec{n}_i\|^2/\|\sqrt{P^{\rm{p}}}\vec{h}_i+\sqrt{N_{\rm{t}}}\vec{n}_i\|^2\right]$. Since the noise $\vec{n}$ is zero mean and Gaussian, we have that $\mathbb{E}\left[\|\sqrt{P^{\rm{p}}}\vec{h}_i^H\vec{h}_i+\sqrt{N_{\rm{t}}}\vec{h}_i^H\vec{n}_i\|^2/\|\sqrt{P^{\rm{p}}}\vec{h}_i+\sqrt{N_{\rm{t}}}\vec{n}_i\|^2\right]=\mathbb{E}\left[\|\sqrt{P^{\rm{p}}}\vec{h}_i^H\vec{h}_i-\sqrt{N_{\rm{t}}}\vec{h}_i^H\vec{n}_i\|^2/\|\sqrt{P^{\rm{p}}}\vec{h}_i-\sqrt{N_{\rm{t}}}\vec{n}_i\|^2\right]$. Thus,
\begin{align*}
g_i(P^{\rm{p}})=\frac{1}{2}\int_{\vec{n}}\left[\underbrace{\frac{\|\sqrt{P^{\rm{p}}}\vec{h}_i^H\vec{h}_i+\sqrt{N_{\rm{t}}}\vec{h}_i^H\vec{n}_i\|^2}{\|\sqrt{P^{\rm{p}}}\vec{h}_i+\sqrt{N_{\rm{t}}}\vec{n}_i\|^2}+\frac{\|\sqrt{P^{\rm{p}}}\vec{h}_i^H\vec{h}_i-\sqrt{N_{\rm{t}}}\vec{h}_i^H\vec{n}_i\|^2}{\|\sqrt{P^{\rm{p}}}\vec{h}_i-\sqrt{N_{\rm{t}}}\vec{n}_i\|^2}}_{\triangleq G_i(P^{\rm{p}})}\right]p_r(\vec{n})\rm{d}\vec{n}\,.
\end{align*}
To prove the concavity of $g_i(P^{\rm{p}})$, we can check the concavity of $G_i(P^{\rm{p}})$. For the sake of simplicity, we denote $A=\|\vec{h}\|^2$, $B=\sqrt{N_{\rm{t}}}\textrm{Re} \{\vec{h}^H\vec{n}\}$, $C=N_{\rm{t}}\|\vec{n}\|^2$ (note that $A$, $B$, and $C$ are different to the ones of Section~\ref{sec:sub:asymptotic}). We also discard the subscript $i$, and further simplify it to
\begin{align*}
G(x)=&\frac{\sqrt{x}A^2 +2\sqrt{x}AB+N_{\rm{t}}\|\vec{h}^H\vec{n}\|^2}{Ax+2\sqrt{x}B+C}+\frac{\sqrt{x}A^2 -2\sqrt{x}AB+N_{\rm{t}}\|\vec{h}^H\vec{n}\|^2}{Ax-2\sqrt{x}B+C}\,.
\end{align*}
 Then, the second derivative of $G(x)$ is
\begin{align*}
G''(x)=\frac{\textrm{d}^2G(x)}{\textrm{d}x^2}=\frac{4(N_{\rm{t}}\|\vec{h}^H\vec{n}\|^2-AC)}{(\|\sqrt{x}\vec{h}_i+\sqrt{N_{\rm{t}}}\vec{n}_i\|^2\|\sqrt{x}\vec{h}_i-\sqrt{N_{\rm{t}}}\vec{n}_i\|^2)^3}G_2(x)\,,
\end{align*}
where $G_2(x)=A^5x^3+3A^4Cx^2+3A^3C^2x-12A^2B^2Cx+A^2C^3-12AB^2C^2+16B^4C$. The denominator of $G''(x)$ is positive. Also, according to the Cauchy–Schwarz inequality, we have that $N_{\rm{t}}\|\vec{h}^H\vec{n}\|^2-AC\leq 0$, and thus $B^2\leq AC$. Then, we know that $G(x)$ is concave when $G_2(x)$ is positive. With $B^2\leq AC$, we have that $G_2(x)\geq A^5x^3+3A^4Cx^2-9A^3C^2x-11A^2C^3+16B^4C$. Let $k\triangleq Ax/C$. We have that $G_2(x)\geq A^2C^3(k^3+3k^2-9k-11)+16B^4C=A^2C^3(k+1)(k+1+2\sqrt{3})(k-(2\sqrt{3}-1))+16B^4C$. Recall that $A\geq 0$, $C\geq 0$. We have that $G_2(x)\geq 0$ when $k\geq 2\sqrt{3}-1$. Recall $C=N_{\rm{t}}\|\vec{n}\|^2$, where $\|\vec{n}\|^2$ follows the Chi-squared distribution with degree of freedom $2N_{\rm{t}}$, and the $99$-th percentile of its lower cumulative distribution is $Q^{-1}(N_{\rm{t}},0,0.99)\sigma_{\rm{n}}^2$. Then, we have that when $P^{\rm{p}}\geq (2\sqrt{3}-1)Q^{-1}(N_{\rm{t}},0,0.99)N_{\rm{t}}\sigma_{\rm{n}}^2/A$, it is almost for sure that $G_i(P^{\rm{p}})$ is concave for all realizations of the noises. Thus, when $P^{\rm{p}}\geq (2\sqrt{3}-1)Q^{-1}(N_{\rm{t}},0,0.99)N_{\rm{t}}\sigma_{\rm{n}}^2/A$, it is almost for sure that $g_i(P^{\rm{p}})$ is concave, which completes the proof.
\end{IEEEproof}
Notice that, the threshold is sufficiently small and is practical for real-world use. For example, let $\sigma^2$ be $-120$~dBm, and $N_{\rm{t}}=16$. We have that $Q^{-1}(N_{\rm{t}},0,0.99)\approx 26.74$. Then, if the power of the received $N_{\rm{t}}$ pilots is larger than approximately $-90$~dBm, then $g(\cdot)$ is concave, i.e., the required transmission power of the pilots is also very low ($P^{\rm{p}}$ should be larger than $-20$~dBm if the pathloss is $70$~dB). Therefore, the requirement is easy to fulfil.

\bibliographystyle{IEEEtran}
\bibliography{sensing3}
\end{document}